\magnification=\magstep1
\input epsf
\voffset=0pt
\vsize=19.8 cm     
\hsize=13.5 cm
\hfuzz=2pt
\tolerance=500
\abovedisplayskip=3 mm plus6pt minus 4pt
\belowdisplayskip=3 mm plus6pt minus 4pt
\abovedisplayshortskip=0mm plus6pt
\belowdisplayshortskip=2 mm plus4pt minus 4pt
\predisplaypenalty=0
\footline={\tenrm\ifodd\pageno\hfil\folio\else\folio\hfil\fi}

\def\la{\mathrel{\hbox{\rlap{\hbox{\lower4pt\hbox{$\sim$}}}\hbox{$<$}}}}
\def\ga{\mathrel{\hbox{\rlap{\hbox{\lower4pt\hbox{$\sim$}}}\hbox{$>$}}}}

\def\arcmin{\hbox{$^\prime$}}
\def\arcsec{\hbox{$^{\prime\prime}$}}
\def\utw{\smash{\rlap{\lower5pt\hbox{$\sim$}}}}
\def\udtw{\smash{\rlap{\lower6pt\hbox{$\approx$}}}}

\def\getsto{\mathrel{\hbox{\rlap{$\gets$}\hbox{\raise2pt\hbox{$\to$}}}}}
\def\lid{\mathrel{\hbox{\rlap{\hbox{\lower4pt\hbox{$=$}}}\hbox{$<$}}}}
\def\gid{\mathrel{\hbox{\rlap{\hbox{\lower4pt\hbox{$=$}}}\hbox{$>$}}}}
\def\sol{\mathrel{\hbox{\rlap{\hbox{\raise4pt\hbox{$\sim$}}}\hbox{$<$}}}
}
\def\sog{\mathrel{\hbox{\rlap{\hbox{\raise4pt\hbox{$\sim$}}}\hbox{$>$}}}
}
\def\lse{\mathrel{\hbox{\rlap{\hbox{\raise4pt\hbox{$<$}}}\hbox{$\simeq$}
}}}
\def\gse{\mathrel{\hbox{\rlap{\hbox{\raise4pt\hbox{$>$}}}\hbox{$\simeq$}
}}}
\def\grole{\mathrel{\hbox{\lower2pt\hbox{$<$}}\kern-8pt
\hbox{\raise2pt\hbox{$>$}}}}
\def\leogr{\mathrel{\hbox{\lower2pt\hbox{$>$}}\kern-8pt
\hbox{\raise2pt\hbox{$<$}}}}
\def\loa{\mathrel{\hbox{\rlap{\hbox{\lower4pt\hbox{$\approx$}}}\hbox{$<$
}}}}
\def\goa{\mathrel{\hbox{\rlap{\hbox{\lower4pt\hbox{$\approx$}}}\hbox{$>$
}}}}

%
%

\font\kleinhalbcurs=cmmib10 scaled 833
\font\eightrm=cmr8
\font\sixrm=cmr6
\font\eighti=cmmi8
\font\sixi=cmmi6
\skewchar\eighti='177 \skewchar\sixi='177
\font\eightsy=cmsy8
\font\sixsy=cmsy6
\skewchar\eightsy='60 \skewchar\sixsy='60
\font\eightbf=cmbx8
\font\sixbf=cmbx6
\font\eighttt=cmtt8
\hyphenchar\eighttt=-1
\font\eightsl=cmsl8
\font\eightit=cmti8

\font\bxf=cmbx10
  \mathchardef\Gamma="0100
  \mathchardef\Delta="0101
  \mathchardef\Theta="0102
  \mathchardef\Lambda="0103
  \mathchardef\Xi="0104
  \mathchardef\Pi="0105
  \mathchardef\Sigma="0106
  \mathchardef\Upsilon="0107
  \mathchardef\Phi="0108
  \mathchardef\Psi="0109
  \mathchardef\Omega="010A
\def\rahmen#1{\vskip#1truecm}
\def\begfig#1cm#2\endfig{\par
\setbox1=\vbox{\rahmen{#1}#2}%
\dimen0=\ht1\advance\dimen0by\dp1\advance\dimen0by5\baselineskip
\advance\dimen0by0.4true cm
\ifdim\dimen0>\vsize\pageinsert\box1\vfill\endinsert
\else
\dimen0=\pagetotal\ifdim\dimen0<\pagegoal
\advance\dimen0by\ht1\advance\dimen0by\dp1\advance\dimen0by1.4true cm
\ifdim\dimen0>\vsize
\topinsert\box1\endinsert
\else\vskip1true cm\box1\vskip4true mm\fi
\else\vskip1true cm\box1\vskip4true mm\fi\fi}
\def\figure#1#2{\smallskip\setbox0=\vbox{\noindent\petit{\bf Fig.\ts#1.\
}\ignorespaces #2\smallskip
\count255=0\global\advance\count255by\prevgraf}%
\ifnum\count255>1\box0\else
\centerline{\petit{\bf Fig.\ts#1.\ }\ignorespaces#2}\smallskip\fi}

\def\xfigure#1#2#3#4{\midinsert\noindent
    $$\epsfxsize=#4truecm\epsffile{#3}$$
    \figure{#1}{#2}\endinsert}


\def\begtab#1cm#2\endtab{\par
\ifvoid\topins\midinsert\vbox{#2\rahmen{#1}}\endinsert
\else\topinsert\vbox{#2\kern#1true cm}\endinsert\fi}
\def\rahmen#1{\vskip#1truecm}
\def\begpet{\vskip6pt\bgroup\petit}
\def\endpet{\vskip6pt\egroup}
\def\begref{\par\bgroup\petit
\let\it=\rm\let\bf=\rm\let\sl=\rm\let\INS=N}
\def\petit{\def\rm{\fam0\eightrm}%
\textfont0=\eightrm \scriptfont0=\sixrm \scriptscriptfont0=\fiverm
 \textfont1=\eighti \scriptfont1=\sixi \scriptscriptfont1=\fivei
 \textfont2=\eightsy \scriptfont2=\sixsy \scriptscriptfont2=\fivesy
 \def\it{\fam\itfam\eightit}%
 \textfont\itfam=\eightit
 \def\sl{\fam\slfam\eightsl}%
 \textfont\slfam=\eightsl
 \def\bf{\fam\bffam\eightbf}%
 \textfont\bffam=\eightbf \scriptfont\bffam=\sixbf
 \scriptscriptfont\bffam=\fivebf
 \def\tt{\fam\ttfam\eighttt}%
 \textfont\ttfam=\eighttt
 \normalbaselineskip=9pt
 \setbox\strutbox=\hbox{\vrule height7pt depth2pt width0pt}%
 \normalbaselines\rm
\def\vec##1{\setbox0=\hbox{$##1$}\hbox{\hbox
to0pt{\copy0\hss}\kern0.45pt\box0}}}%
\let\ts=\thinspace
%
\font \tafontt=     cmbx10 scaled\magstep2
\font \tafonts=     cmbx7  scaled\magstep2
\font \tafontss=     cmbx5  scaled\magstep2
\font \tamt= cmmib10 scaled\magstep2
\font \tams= cmmib10 scaled\magstep1
\font \tamss= cmmib10
\font \tast= cmsy10 scaled\magstep2
\font \tass= cmsy7  scaled\magstep2
\font \tasss= cmsy5  scaled\magstep2
\font \tasyt= cmex10 scaled\magstep2
\font \tasys= cmex10 scaled\magstep1
\font \tbfontt=     cmbx10 scaled\magstep1
\font \tbfonts=     cmbx7  scaled\magstep1
\font \tbfontss=     cmbx5  scaled\magstep1
\font \tbst= cmsy10 scaled\magstep1
\font \tbss= cmsy7  scaled\magstep1
\font \tbsss= cmsy5  scaled\magstep1

\newbox\chsta\newbox\chstb\newbox\chstc
\def\centerpar#1{{\advance\hsize by-2\parindent
\rightskip=0pt plus 4em
\leftskip=0pt plus 4em
\parindent=0pt\setbox\chsta=\vbox{#1}%
\global\setbox\chstb=\vbox{\unvbox\chsta
\setbox\chstc=\lastbox
\line{\hfill\unhbox\chstc\unskip\unskip\unpenalty\hfill}}}%
\leftline{\kern\parindent\box\chstb}}
 \def \chap#1{
    \vskip24pt plus 6pt minus 4pt
    \bgroup
 \textfont0=\tafontt \scriptfont0=\tafonts \scriptscriptfont0=\tafontss
 \textfont1=\tamt \scriptfont1=\tams \scriptscriptfont1=\tamss
 \textfont2=\tast \scriptfont2=\tass \scriptscriptfont2=\tasss
 \textfont3=\tasyt \scriptfont3=\tasys \scriptscriptfont3=\tenex
     \baselineskip=18pt
     \lineskip=18pt
     \raggedright
     \pretolerance=10000
     \noindent
     \tafontt
     \ignorespaces#1\vskip7true mm plus6pt minus 4pt
     \egroup\noindent\ignorespaces}%
 \def \sec#1{
     \vskip25true pt plus4pt minus4pt
     \bgroup
 \textfont0=\tbfontt \scriptfont0=\tbfonts \scriptscriptfont0=\tbfontss
 \textfont1=\tams \scriptfont1=\tamss \scriptscriptfont1=\kleinhalbcurs
 \textfont2=\tbst \scriptfont2=\tbss \scriptscriptfont2=\tbsss
 \textfont3=\tasys \scriptfont3=\tenex \scriptscriptfont3=\tenex
     \baselineskip=16pt
     \lineskip=16pt
     \raggedright
     \pretolerance=10000
     \noindent
     \tbfontt
     \ignorespaces #1
     \vskip12true pt plus4pt minus4pt\egroup\noindent\ignorespaces}%
 \def \subs#1{
     \vskip15true pt plus 4pt minus4pt
     \bgroup
     \bxf
     \noindent
     \raggedright
     \pretolerance=10000
     \ignorespaces #1
     \vskip6true pt plus4pt minus4pt\egroup
     \noindent\ignorespaces}%
 \def \subsubs#1{
     \vskip15true pt plus 4pt minus 4pt
     \bgroup
     \bf
     \noindent
     \ignorespaces #1\unskip.\ \egroup
     \ignorespaces}
\def\footnoterule{\kern-3pt\hrule width 2true cm\kern2.6pt}
\newcount\footcount \footcount=0
\def\advftncnt{\advance\footcount by1\global\footcount=\footcount}
\def\fonote#1{\advftncnt$^{\the\footcount}$\begingroup\petit
       \def\textindent##1{\hang\noindent\hbox
       to\parindent{##1\hss}\ignorespaces}%
\vfootnote{$^{\the\footcount}$}{#1}\endgroup}

\newcount\sterne
\outer\def\byebye{\bigskip\typeset
\sterne=1\ifx\speciali\undefined\else
\bigskip Special caracters created by the author
\loop\smallskip\noindent special character No\number\sterne:
\csname special\romannumeral\sterne\endcsname
\advance\sterne by 1\global\sterne=\sterne
\ifnum\sterne<11\repeat\fi
\vfill\supereject\end}
\def\typeset{\centerline{\petit This article was processed by the author
using the \TeX\ Macropackage from Springer-Verlag.}}
 
\def\eck#1{\left\lbrack #1 \right\rbrack}
\def\eckk#1{\bigl[ #1 \bigr]}
\def\rund#1{\left( #1 \right)}
\def\abs#1{\left\vert #1 \right\vert}
\def\wave#1{\left\lbrace #1 \right\rbrace}
\def\ave#1{\left\langle #1 \right\rangle}

\def\part#1#2{{\partial #1\over\partial #2}}

\def\arcsecf {\hbox{$.\!\!^{\prime\prime}$}}
\def\arcminf {\hbox{$.\!\!^{\prime}$}}
{\catcode`\@=11
\gdef\SchlangeUnter#1#2{\lower2pt\vbox{\baselineskip 0pt \lineskip0pt
  \ialign{$\m@th#1\hfil##\hfil$\crcr#2\crcr\sim\crcr}}}
}
\def\gtrsim{\mathrel{\mathpalette\SchlangeUnter>}}
\def\lesssim{\mathrel{\mathpalette\SchlangeUnter<}}

\def\i{{\rm i}}

\def\Real{{\rm I\mathchoice{\kern-0.70mm}{\kern-0.70mm}{\kern-0.65mm}%
  {\kern-0.50mm}R}}
\def\C{\rm C\kern-.42em\vrule width.03em height.58em depth-.02em
       \kern.4em}
\font \bolditalics = cmmib10
\def \vc #1{{\textfont1=\bolditalics \hbox{$\bf#1$}}}
{\catcode`\@=11
\def\malen#1#2{\global \advance\fnum by 1\midinsert\vskip0.5truecm\noindent%
$${\epsfxsize=0.5\hsize\epsffile{#1}}$$
     $$\vbox{\hsize=158truemm{
        {\eightpoint\par\noindent
    {\bf Figure \the\fnum:~}#2}}}$$\medskip\noindent\endinsert}
\def\ma#1#2{\global \advance\fnum by 1\vskip0.5truecm\noindent%
$${\epsfxsize=0.6\hsize\epsffile{#1}}$$
     $$\vbox{\hsize=158truemm{
        {\eightpoint\par\noindent
    {\bf Figure \the\fnum:~}#2}}}$$\medskip\noindent}

\def\eps{{\epsilon}}

\def\SFB{{This work was supported by the ``Sonderforschungsbereich
375-95 f\"ur
Astro--Teil\-chen\-phy\-sik" der Deutschen For\-schungs\-ge\-mein\-schaft.}}

\chap{The Mass Distribution of CL0939+4713 obtained from a `Weak' Lensing
Analysis of a WFPC2 image}

{\bf Carolin Seitz$^1$, Jean-Paul Kneib$^2$, Peter Schneider$^1$ \& Stella Seitz$^1$}
\bigskip
$^1$ Max-Planck-Institut f\"ur Astrophysik, Postfach 1523, D-85740
Garching, Germany.

$^2$ Institute of Astronomy, Madingley Road, Cambridge CB3 0HA, UK.

\sec{Abstract}
The image distortions of high-redshift galaxies caused by
gravitational light deflection of foreground clusters of galaxies can
be used to reconstruct the two-dimensional surface mass density of
these clusters.  We apply an unbiased parameter-free reconstruction
technique to the cluster CL0939+4713 (Abell 851), observed with the
WFPC2 on board of the HST. We demonstrate that a single deep WFPC2
observation can be used for cluster mass reconstruction despite its
small field of view and the irregular shape of the data field
(especially for distant clusters).  For CL0939, we find a strong
correlation between the reconstructed mass distribution and the bright
cluster galaxies indicating that mass follows light on average. The
detected anti-correlation between the faint galaxies and the
reconstructed mass is most likely an effect of the magnification
(anti) bias, which was detected previously in the cluster
A1689. Because of the high redshift of CL0939 ($z_d=0.41$), the
redshift distribution of the lensed, faint galaxies has to be
accounted for in the reconstruction technique.  We derive an
approximate global transformation for the surface mass density which
leaves the mean image ellipticities invariant, resulting in an
uncertainty in the normalization of the mass.  From the non-negativity
of the surface mass density, we derive lower limits on the mass inside
the observed field of $0.75({h^{-1}_{50}\hbox{\ts \rm Mpc}})^2$
ranging from $M>3.6 \times 10^{14}h^{-1}_{50} M_\odot$ to $ M>6.3
\times 10^{14}h^{-1}_{50} M_\odot$ for a mean redshift of $\ave z=1$
to $\ave z=0.6$ of the faint galaxy images with
$R\in(23,25.5)$. However, we can break the invariance transformation
for the mass using the magnification effect on the observed number
density of the background galaxies. Assuming a mean redshift of $\ave
z=0.8$ and a fraction of $x=15\%$ ($x=20\%$) of cluster galaxies in
the observed galaxy sample with $R\in(23,25.5)$ we obtain for the mass
inside the field $M\approx 5 \times 10^{14}h^{-1}_{50} M_\odot$
($M\approx 7 \times 10^{14}h^{-1}_{50} M_\odot$) which corresponds to
$M/L\approx 100 h_{50}$ ($M/L\approx 140 h_{50}$).

\vfill \eject

\sec{1. Introduction}
Since the pioneering work of Tyson, Valdes and Wenk (1990) it has been
realized that the weak shearing effects of clusters introduced on
images of faint background galaxies can be used to obtain the mass
distribution of these lensing clusters (for a recent review on cluster
lensing, see Fort \& Mellier 1994; see also Kochanek 1990;
Miralda-Escud\'e 1991). 
Kaiser \& Squires (1993, hereafter KS) have
derived an explicit expression for the two-dimensional surface mass
density as a function of the shear (or tidal gravitational field)
caused by the cluster, which in turn can be obtained from the
distorted images of background galaxies. This inversion method has
been applied to several clusters observed from the ground (Fahlman et
al.\ts 1994, Smail et al.\ts 1995a, Kaiser at al.\ts 1995),
demonstrating the applicability of this new method to determine mass
profiles and total mass estimates of clusters. The detection of
sheared images far out in the cluster 0024+16 (Bonnet, Mellier \& Fort
1994; Kassiola et al.\ts 1994) shows that weak lensing can investigate
previously unexplored regions in clusters.

Recently, the KS inversion technique has been modified
and generalized to account for strong lensing, as it should occur near
the center of clusters (Schneider \& Seitz 1995, Seitz \& Schneider
1995a, Kaiser 1995), and for the finite data field defined by the CCD size
(Schneider 1995, Kaiser et al.\ts 1995, Bartelmann 1995,
Seitz \& Schneider 1995b, henceforth SS). In 
SS, a detailed quantitative comparison between the various
inversion techniques has been made, and it was demonstrated that the
inversion formula derived in SS is the most accurate of the unbiased
ones. In particular, if the cluster mass distribution is significantly
more extended than the data field (i.e. the CCD), the SS inversion
formula is significantly more accurate than the other currently known
inversion techniques. 

Such a situation generally occurs if the data are taken with the WFPC2
on board the {\sl Hubble Space Telescope} (HST), owing to its fairly
small field of view. Hence, if WFPC2 images are used for the
reconstruction of the surface mass density of a cluster, it is
necessary to use a finite-field inversion formula such as the one
derived in SS.  As was pointed out in Schneider \& Seitz 1995, even
then the mass density can be derived only up to a global invariance
transformation, which is the mass-sheet degeneracy found by Gorenstein
et al.\ts (1988). The invariance transformation may be broken if the
magnification effects are taken into account which changes the local
number density of images of an appropriately chosen subset of faint
galaxies (Broadhurst, Taylor \& Peacock 1995), and which changes the
size of galaxy images at fixed surface brightness -- which is unchanged
by gravitational light deflection (Bartelmann \& Narayan 1995). In
particular, this latter paper demonstrates that the inclusion of
magnification effects may improve the cluster mass inversion
considerably, and can also provide a unique means to determine the
 redshift distribution down to very faint magnitudes.

In this paper we present the first application of a finite-field
cluster inversion to the deep WFPC2 observation (10 orbits)
 of the distant cluster CL 0939+4713 retrieved from the HST archive.
These data have been used for the study of the morphology of the
cluster galaxies and the Butcher--Oemler effect by Dressler et al.
(1994a). In Sect.\ts 2 we briefly describe the data, and discuss the
image identification and the determination of the image shapes, which
is used for the estimate of the local image distortion. Sect.\ts 3
briefly summarized the inversion method, the results of which are
presented in Sect.\ts 4 and discussed in Sect.\ts 5. 

\sec{2. Observation and data analysis}
\begfig 13 cm
\figure{1a}
{Observations of the cluster CL0939+4713 obtained with WFPC2 using the
$702W$ filter and an exposure time of $21000 s$.  The side-length of the
data field is $2\arcminf 5$ ($1h^{-1}_{50}$\ts Mpc) for an EdS-universe
with $H_0=50h_{50}$\ts km/s/Mpc, 1 arcsec on the sky represents $6.51
h_{50}^{-1}$\ts kpc).
Dressler \& Gunn (1992) propose the cluster center to be close to the 3
bright galaxies in the upper left corner of the lower left CCD. North
is at the bottom east to the right}
\endfig

Cl0939+4713 was observed in January 1994 with the WFPC2 camera on the
Hubble Space Telescope (Dressler et al. 1994a).  The observation
consists of 10 single
orbits of 2100 seconds (or a total exposure time of 5h50min) and
corresponds probably to the deepest cluster observation done with the
HST/WFPC2.  The exposures were divided into two groups of 5 with a shift
of 10 pixel East and 20 pixels South between the two groups.  After
StSci pipeline processing, the data were shifted and combined to remove
cosmic rays and hot pixel using standard STSDAS/IRAF routines.  A
mosaic of the 3 WFC chips and the PC chips was constructed, though due
to the smaller pixel size of the PC chip and therefore a brighter
isophotal limit we discard it from the analysis.  The image was then
run through the SExtractor package (Bertin \& Arnouts 1995) to detect
objects, measure their magnitudes, mean isophotal surface brightness
and second moments.  All objects with isophotal areas larger than 12
pixels and higher than 2$\sigma$/pixel ($\mu_{F702W}=25.3 {\rm
mag/arcsec^{2}}$) were selected.  For each object the unweighted first
and second moments were computed to determine their center, their size,
their ellipticity and orientation.  To convert instrumental F702W
magnitudes into standard $R$ we use the synthetic zero point and color
corrections listed in Holtzman et al.\ (1995). For the color term we
choose $(V-R) \simeq0.6$ typical of a $z \sim 0.8$ late-type spiral.
The color correction is then $+0.2$ mag, and remains small for other
choices of the
colour term.  The typical photometric errors of our faintest objects,
$R<26.5$, are $\delta R \sim $0.1--0.2.  A neural network algorithm was
used to identify stellar objects, $22$ of those were detected.  A
galaxy catalogue was then constructed with a total of $572$ galaxies
down to R=26.5.

Fig.\ts 1a shows the full WFPC2 image of the cluster, and Fig.\ts 1b a
zoomed image of the region marked in Fig.\ts 1a. A detailed inspection
discovered the arc candidate and a likely pair in this central cluster
region. These strong lensing features confirm that the cluster is
over-critical and probably indicate the densest part of the cluster. The
spectroscopic observation of the bright pair (R=22.5, and R=22.9 for
it's counter-image candidate) will confirm or otherwise the lensing
assumption.  If it is indeed a gravitational pair, it will constrain
strongly the mass distribution of the very central part of the
cluster.


\begfig 12 cm
\figure{1b}
{A zoomed image of the region marked in Fig.\ts 1a. We find an
arc-candidate A0 and a likely gravitational pair P1\& 2 with the
counter image P3}
\endfig

Fig.\ts 2 shows the number vs. magnitude diagram of the galaxies
detected in the field (solid line). These numbers are compared with the
field galaxies counts in the R band from Smail et al (1995b).
It is clear that most of the galaxies with $R\in(19,22)$
are likely cluster members. Furthermore there is a likely contamination
from cluster members of $\sim$ 150 objects down to $R=25.5$ within the WFC
field; the cluster contamination is expected to be higher in the central
part than in the outer part.

\xfigure{2}{The number vs. magnitude diagram of all galaxies detected
within the WFC field (solid line). The dotted line shows the number
counts -- rescaled to the area of WFC field -- from Smail et al.\ts
(1995b), which yields $N(R)\propto 10^{\gamma R}$, with
$\gamma=0.32$ and normalization such that $N(R<27)=7.3 \times 10^5$
per square degree.  Assuming that the dotted line represents the
counts of the faint galaxies, the dashed-dotted histogram gives the
number counts versus magnitude of the cluster galaxies}{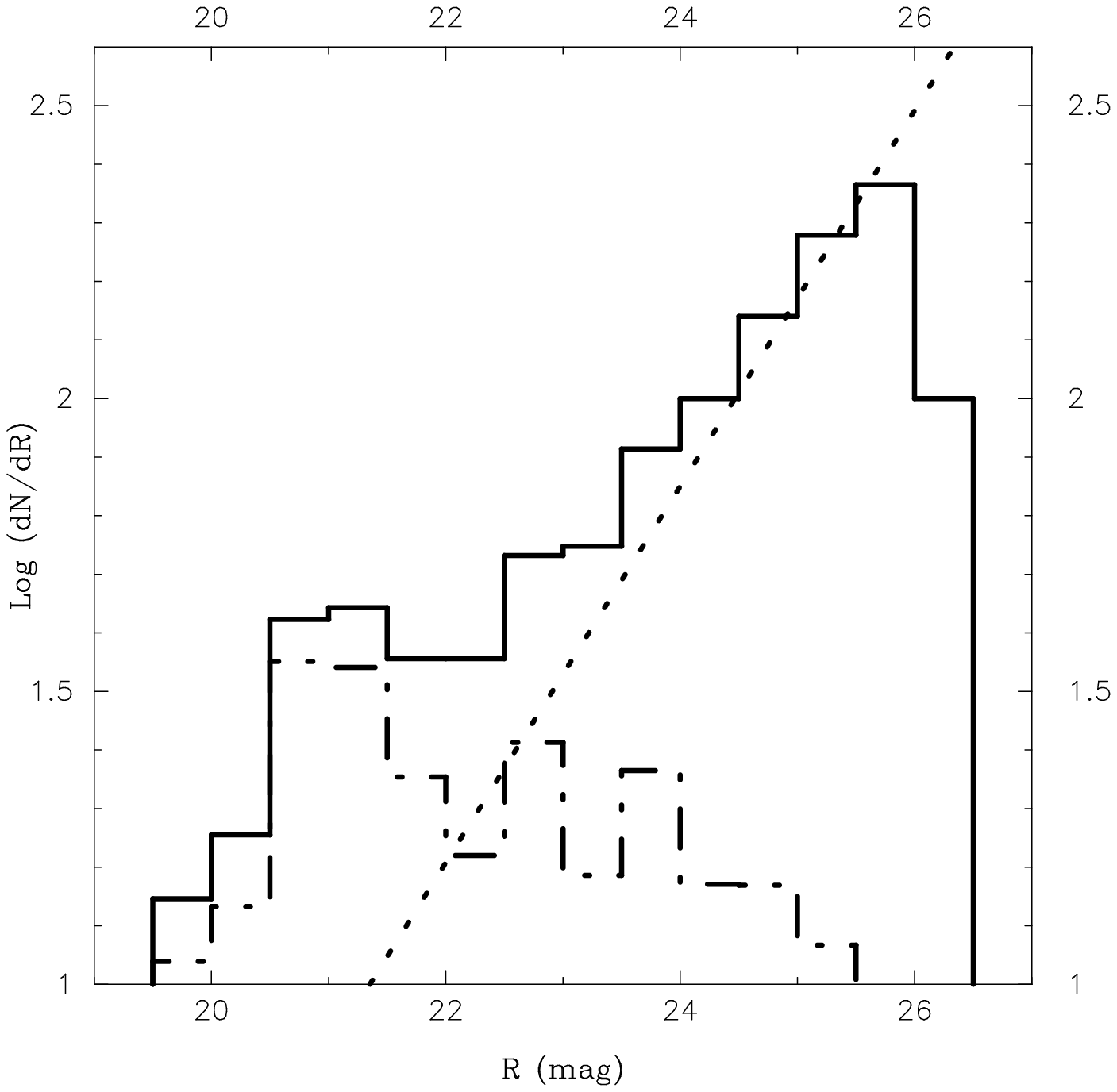}{8}

From the second moments of light
$Q_{ij}$  we calculate for each galaxy image the (complex) ellipticity
$$
\chi={Q_{11}-Q_{22}+2{\rm i}Q_{12}\over Q_{11}+Q_{22}}=\abs{\chi} e^{2\i\vartheta}\ .	
\eqno(2.1a)
$$
However, it is more convenient to work in terms of the ellipticity
parameter $\eps$ which has the same phase $\theta$ as $\chi$, and
modulus
$$
\abs{\eps}={1-r\over 1+r}\quad{\rm with}\quad r=\sqrt{1-\abs \chi\over
1+\abs \chi} \ .
\eqno (2.1b)
$$
Then we use these image ellipticities $\eps_k$ of a galaxy at
$\vc\theta_k$ to calculate 
the local mean image ellipticity on a grid $\vc \theta_{ij}$  
$$
\bar \eps(\vc \theta_{ij})={\sum_{k=1}^{N_{\rm gal}}\eps_k
\; u\rund{\abs{\vc\theta_{ij}-\vc\theta_k}}
\over \sum_{k=1}^{N_{\rm gal}}u\rund{\abs{\vc\theta_{ij}-\vc\theta_k}}}
\ , \eqno(2.2)
$$
with the weight factor
$$
u(d)=\exp(-d^2/s^2)\; ,
$$
with a smoothing length $s$, which, unless noted otherwise, is chosen
as $s=0\arcminf3$ (117 $h^{-1}_{50}$ kpc).  The resulting map of $\bar
\eps(\vc \theta)$ is shown in Fig.\ts 3 using all images covering more
than $12$ pixels (pixel-size $0\arcsecf1$) and using four different
magnitude cuts for the galaxies:  $R\in(24,25.5)$ for the upper left
panel, $R\in(23,25.5)$ for the upper right, $R\in(22,25.5)$ for the
lower left and $R\in (21,25.5)$ for the lower right. The corresponding
numbers of galaxies used for constructing the shear maps are 226, 295
343, and 383, respectively, meaning that the average number of
galaxies having a distance of less than the smoothing length
from the point $\vc \theta_{ij}$ is about 13, 17, 20 and 22.
The cut at fainter
magnitude was chosen in order to be not too much contaminated by the
circularization effect of measuring small galaxies with poor signal to
noise. We find that the ``shear field'' is quite robust under adding
brighter galaxies to the sample: the direction of the local shear
vector is almost unchanged and its absolute value is decreased on
average for the brighter galaxy samples, especially in regions close
to the cluster center. The reason for this is that the modulus of the
expectation value of the mean image ellipticities is smaller for
background galaxies closer to the cluster and it is zero for cluster-
and foreground galaxies, both leading to a decrease in the mean image
ellipticities for the brighter galaxy samples.  The direction of the
expectation value of the mean image ellipticities is not changed, since
the mean image orientation of the background galaxies does not depend
on their distances to the cluster, and since cluster- and foreground
galaxies show no preferred alignment at all.

\xfigure{3}{The orientation and absolute value of the local mean image
ellipticities $\bar \eps=\abs{\bar \eps}\rund{\cos 2\varphi+i
\sin 2\varphi}$ of
galaxies with $R\in(24,25.5)$ (upper left), $R\in(23,25.5)$ (upper
right), $R\in(22,25.5)$ (lower left) and $R\in(21,25.5)$ (lower
right). We exclude galaxies covering less than $12$ pixels (pixel-size
$0\arcsecf1$). We choose a smoothing length of $s=0\arcminf3$; as is
clearly seen, the `coherence length' of the shear pattern is larger
than this smoothing length. The
vectors displayed include an angle of $\varphi$ with the x-axis, and a
mean image ellipticity $\abs{\bar \eps}=1$ would correspond to a
vector of length $0\arcminf4$}
{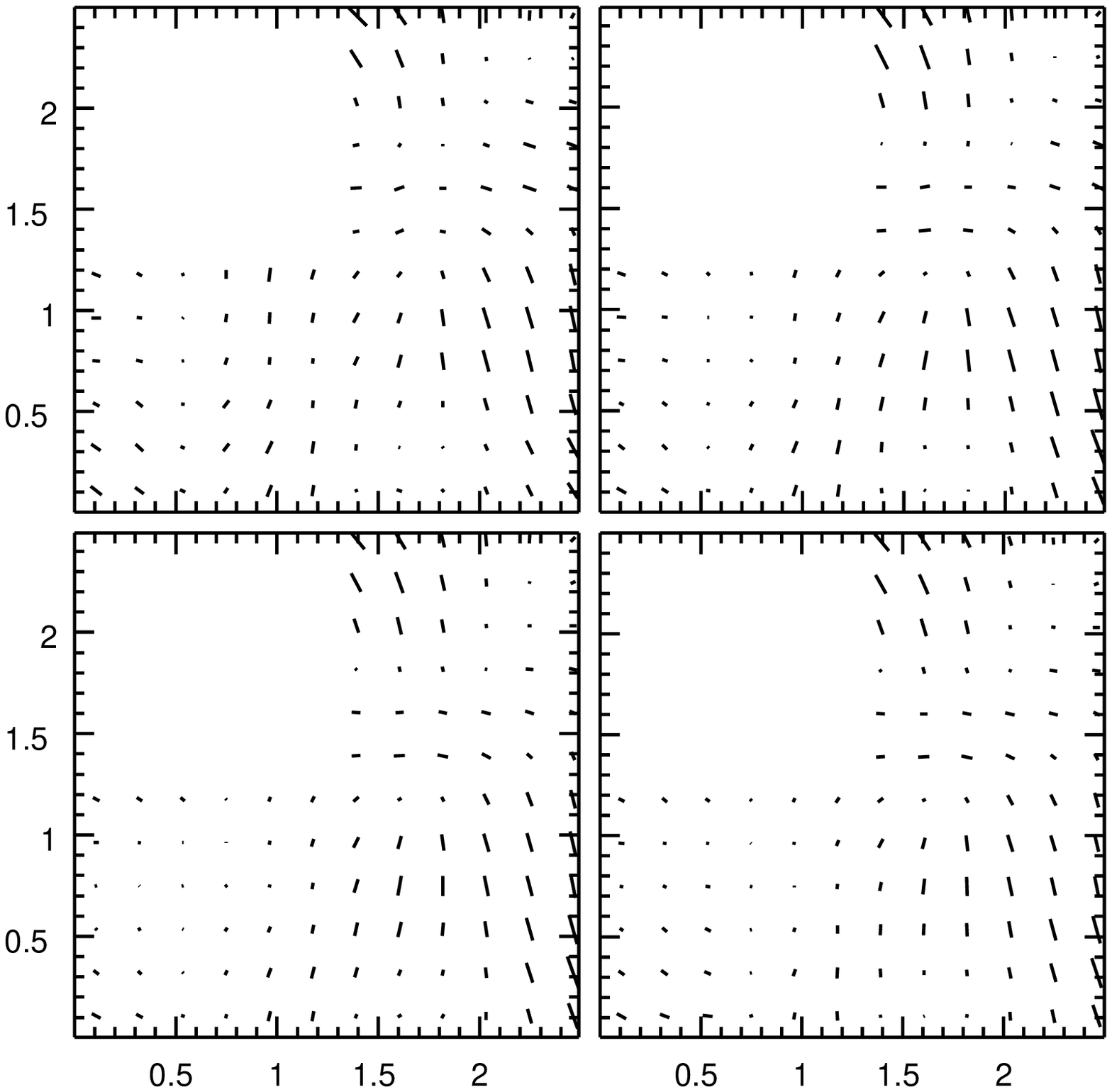}{10}

\sec{3. Method of Reconstruction}
In this section we briefly describe the reconstruction of the cluster
surface mass density using the observed
 map of mean image ellipticities.  Due to the high redshift of the
cluster ($z_d=0.41$) we can not assume that all source galaxies are at
the same effective distance to the cluster, i.e. that their
$D(z_d,z)/D(z)$ is the same.  Therefore, we relate the critical surface
mass density
$$
\Sigma_{\rm crit}(z)={c^2D(z)\over 4\pi G D(z_d)D(z_d,z)}
$$ 
for a source at redshift $z$ to the critical surface mass density
for a source at `infinity' by defining $w(z)$ through
$$
w(z) \; \Sigma_{\rm crit}(z)= \lim_{z\to \infty}\Sigma_{\rm crit}(z)\ ,
$$
for $z>z_d$, and $w(z)=0$ for $z\le z_d$,
and obtain for the dimensionless surface mass density $\kappa(\vc\theta,z)$ and
the shear $\gamma(\vc\theta,z)$
$$
\kappa(\vc\theta,z)=w(z) \kappa_\infty(\vc\theta)\quad , \quad
\gamma(\vc\theta,z)=w(z) \gamma_\infty(\vc\theta)   \quad .
\eqno(3.1)
$$
The form of $w(z)$ depends on the geometry of the universe, and for 
an Einstein-de Sitter universe we have
$$
w(z)=\cases
{\displaystyle\;  0 \qquad \qquad \quad \qquad \qquad \hbox{for } \quad z \le z_d ; \cr
\displaystyle {\sqrt{1+z}-\sqrt{1+z_d}\over \sqrt{1+z}-1} \qquad
\hbox{for } \quad z > z_d . \cr}
\eqno(3.2)
$$

The following description of the reconstruction of the surface mass
density is based on the simplifying assumption that the cluster is not
critical, i.e., $(1-w\kappa_\infty)^2-w^2\gamma_\infty^2>0$ for all
sources. However, we point out that all the resulting mass maps shown
in this paper have been calculated {\it without} this assumption,
using the more complicated method described in Seitz \& Schneider
(1995c). However, since the reconstruction is much easier to describe
for non-critical clusters, we describe here the reconstruction of
non-critical clusters only. We also note that there are only minor
changes in the results if this assumption is introduced.

As described in Seitz \& Schneider (1995c), the local 
expectation value of the image ellipticity can be approximated through
$$
\ave \eps\approx {-\ave w \gamma_\infty\over 1-f \ave w \kappa_\infty}
\ ,
\eqno(3.3)
$$
if $\kappa_\infty \lesssim 0.8$ and if the mean redshift of the
sources is $\ave z \gtrsim 0.7$ for this particular cluster redshift.
In (3.3) we used the definitions
$$
\ave{w^k}=\int_0^\infty dz \; p_s(z)\; w^k(z) 
\qquad {\rm and} \qquad  f={\ave {w^2}\over \ave w^2}\ ,
\eqno(3.4)
$$
where  $p_s(z)$ is the redshift distribution of the sources.
>From (3.3) we find that the transformation
$\kappa_\infty\to \kappa_\infty '$ with
$$
\lambda\left (1-\kappa_\infty {\ave{ w^2}\over \ave w} \right)=
       \left( 1-\kappa'_\infty{\ave{ w^2}\over \ave w}\right) \,
\eqno(3.5)
$$
which implies that $\gamma'_\infty =\lambda \gamma_\infty$,
leaves the image ellipticities
unchanged. Therefore, using only image ellipticities for the
reconstruction, $1-f \ave w\kappa_\infty$ can be derived only up
to a multiplicative constant.
Using the relation between the gradient of the surface mass density
and the derivatives of the shear (Kaiser 1995), we obtain from (3.3)
with $K(\vc\theta):=\ln [ 1-f \ave w \kappa_\infty(\vc\theta) ]$
$$
\vc \nabla {  K}
={1\over 1-f^2 \abs{\ave{ \eps}}^2}
\left ( \matrix{ 1-f \ave{\eps_1}  & -f\ave { \eps_2} \cr   
-f\ave{ \eps_2 } &
1+f\ave{ \eps_1}  \cr } \right )
\left ( \matrix{ -f\ave{\eps_1}_1  -f\ave{ \eps_2}_2  \cr
-f\ave{\eps_2}_1 +f\ave{\eps_1}_2  \cr } \right )=: \vc u(\vc\theta)\ ,
\eqno(3.6)
$$
with $\ave \eps=\ave{\eps_1}+{\rm i}\ave{\eps_2}$ and gradients
$\ave{\eps_i}_j=\partial\ave{\eps_i} /\partial \theta_j$. 
Since the mean image ellipticity $\bar\eps$ provides an unbiased
estimator of the expectation value $\ave{\eps}$, we set
$\ave{\eps}\approx \bar\eps$. Then $\vc u(\vc\theta)$ can be
determined from observations, for an assumed value of $f$,
which characterizes the redshift distribution of the sources. From
that, $K(\vc\theta)$ is obtained as 
$$
K(\vc \theta )=\int_{\cal U}\; d^2\theta' \; \vc H(\vc
\theta,\vc\theta')\cdot  \vc u(\vc \theta') +\bar K \ .
\eqno(3.7)
$$
In Eq.(3.7), the kernel $\vc H(\vc \theta,\vc \theta')$ is calculated for
the data field $\cal U$ according to the method suggested by Seitz \&
Schneider (1995b).
So far we have not calculated the kernel $\vc H$ for the irregularly-shaped 
WFPC2 field. Therefore, we reconstruct $K$ on two rectangular fields with
side-length of about $2\arcminf5\times 1\arcminf25$ and
$1\arcminf25\times 2\arcminf5$. Since we have an additive constant free, we
shift one of the resulting $K$-maps such that the mean of $K$ inside
the overlapping region of the two data fields is the same. Then, the
resulting mass map is obtained by joining together these two
independent reconstructions at the diagonal of the lower right
CCD. That means that all mass maps shown here display a discontinuity
at this diagonal; however, the jump across this line is always
remarkably small,
indicating the relative uncertainty of the reconstruction.

The redshift distribution of the field galaxies down to the faint
magnitude limits considered here is poorly
known. Redshift surveys of considerably brighter galaxies indicate
that the redshift distribution is fairly broad, and a high-redshift
tail cannot be excluded (see Lilly 1993, Colless et al.\ts 1993, and
Cowie et al.\ts 1995,
and references therein).
We therefore take the same parameterization of $p_s(z)$ as used in
Brainerd, Blandford \& Smail (1995),
$$
p_s(z)={\beta z^2\over \Gamma
\rund{3/\beta}z_0^3}\exp\rund{-\rund{z/z_0}^\beta} \ .
\eqno(3.8)
$$

We consider different values of the parameter $\beta$ for which the mean
redshift is given through $\ave
z={z_0 \Gamma(4/\beta)/ \Gamma(3/\beta)}$. In Fig.\ts 4 we show the
distribution $p_s(z)$ for $\beta=1,1.5,3$ and 
$\ave z\in\wave{0.8,1.0,1.5}
$ (right panels) and the moments $\ave w,
\ave{w^2}$ and $f=\ave{w^2}/\ave w^2$ as a function of the mean
redshift $\ave z$ for the redshift $z_d=0.41$ of CL 0939+4713 
(left panels). Kneib et al.\ts (1995) attempted using the Abell 2218 cluster-lens
to determine the mean redshift distribution of the faint galaxies and found
a mean value of $<z>\sim 0.8$ for $23.5<R<25.5$ which is above but
consistent with the non-evolution expectation. Therefore, our choice
of the  parametrization
of the redshift distribution shall be close to the true distribution of
faint galaxies.

\xfigure{4}{The redshift distribution $p_s(z)$, defined in
Eq.\ts(3.8), is shown in the right panels for a mean redshift of $\ave
z=0.8$ (top) $\ave z=1 $ (middle) and $\ave z=1.5$ (bottom), and for
the parameters $\beta=1$ (solid line), $ \beta=1.5 $ (dotted line) and
$\beta=3$ (dashed line).  The left panels show the moments $\ave w$,
$\ave{w^2}$ and the ratio $f=\ave{w^2}/\ave w^2$ defined in
Eq.\ts(3.4) as a function of mean redshift $\ave z$ for a cluster
redshift $z_d=0.4$} {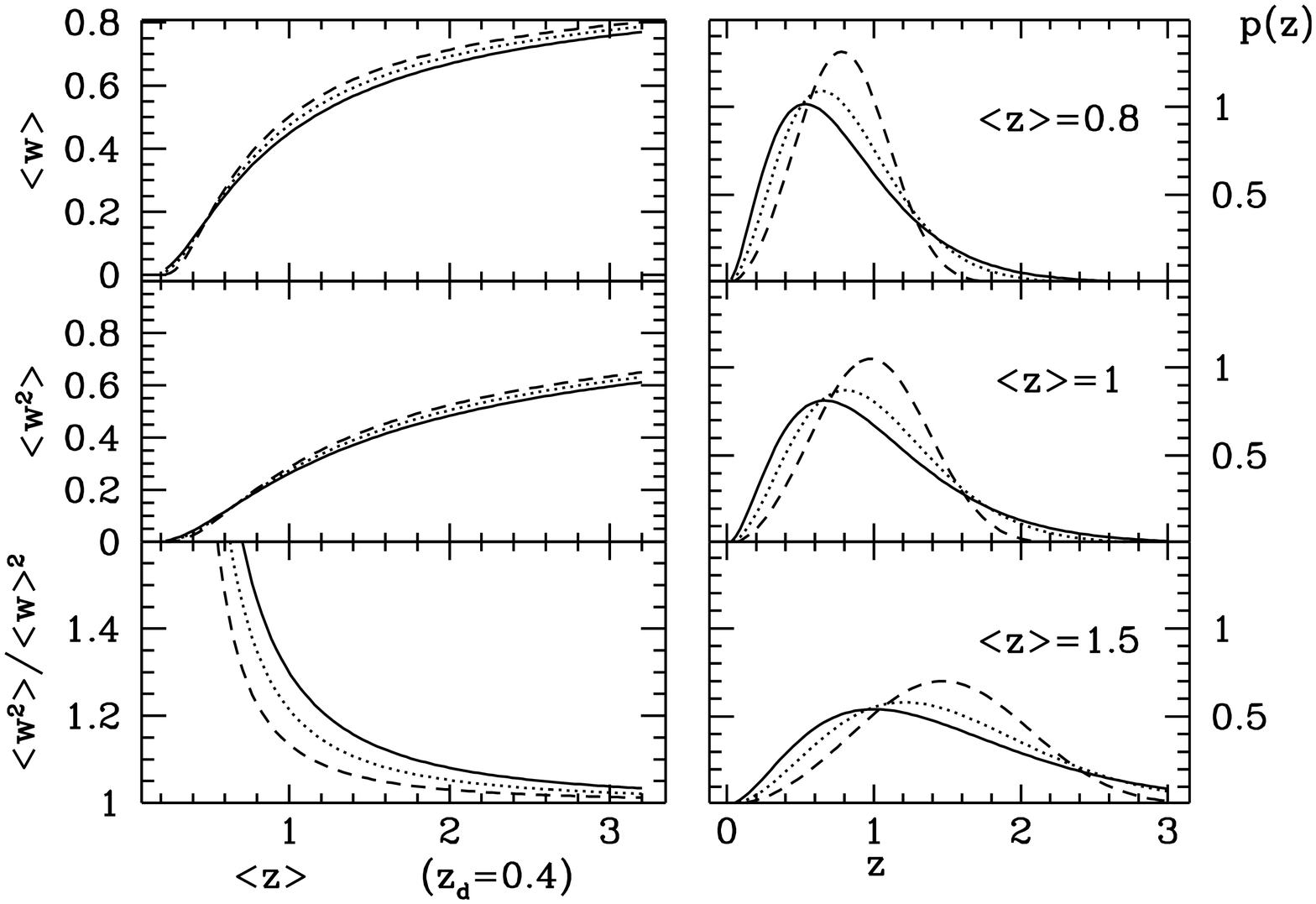}{12}

\sec{4. Results}

\subs{4.1 The reconstructed mass distribution}

In Fig.5 we show the reconstructed surface mass density for different
mean redshifts $\ave z$, using in each case galaxies with
$R\in(23,25.5)$ and the invariance transformation (3.5) such that the
minimum of the resulting $\kappa_\infty$-map is roughly zero to avoid
unphysical negative surface mass densities. We see that for a mean
redshift of about $\ave z=0.6-0.8$ of the faint galaxies in this
magnitude interval, the cluster
is quite strong and could indeed be (marginally) critical. We identify
four main features, i.e., the two local maxima (the `first' in the
lower left quadrant of the field, and the `second' at the boundary
between the lower left and lower right quadrant), the overall increase
of $\kappa_\infty$ towards the first maximum in the lower two
quadrants, and a minimum in the upper right quadrant.

\xfigure{5}
{The reconstructed surface mass density of the cluster CL
0939+4713. For the reconstruction we use 295 galaxy images with $R\in
(23,25.5)$ and assume that their redshift distribution is given
through (3.8) with $\beta=1$ and $\ave z=0.6$ (upper left), $\ave
z=0.8$ (upper right), $\ave z=1$ (lower left) or $\ave z=1.5$ (lower
right). For all these reconstructions we use a smoothing length of
$s=0\arcminf 3$ in the weight function appearing in Eq.(2.2). Since we
use no data on the upper left quadrant, and therefore cannot
reconstruct the surface mass density there, we arbitrarily set
$\kappa=0$ in this quadrant; this leads to the `funny' shape in the
level plots and the jumps in the corresponding contour plots, which
are seen throughout this paper. The small discontinuity along the
diagonal of the lower right quadrant is due to joining together two
independent reconstructions, as described in the text}
{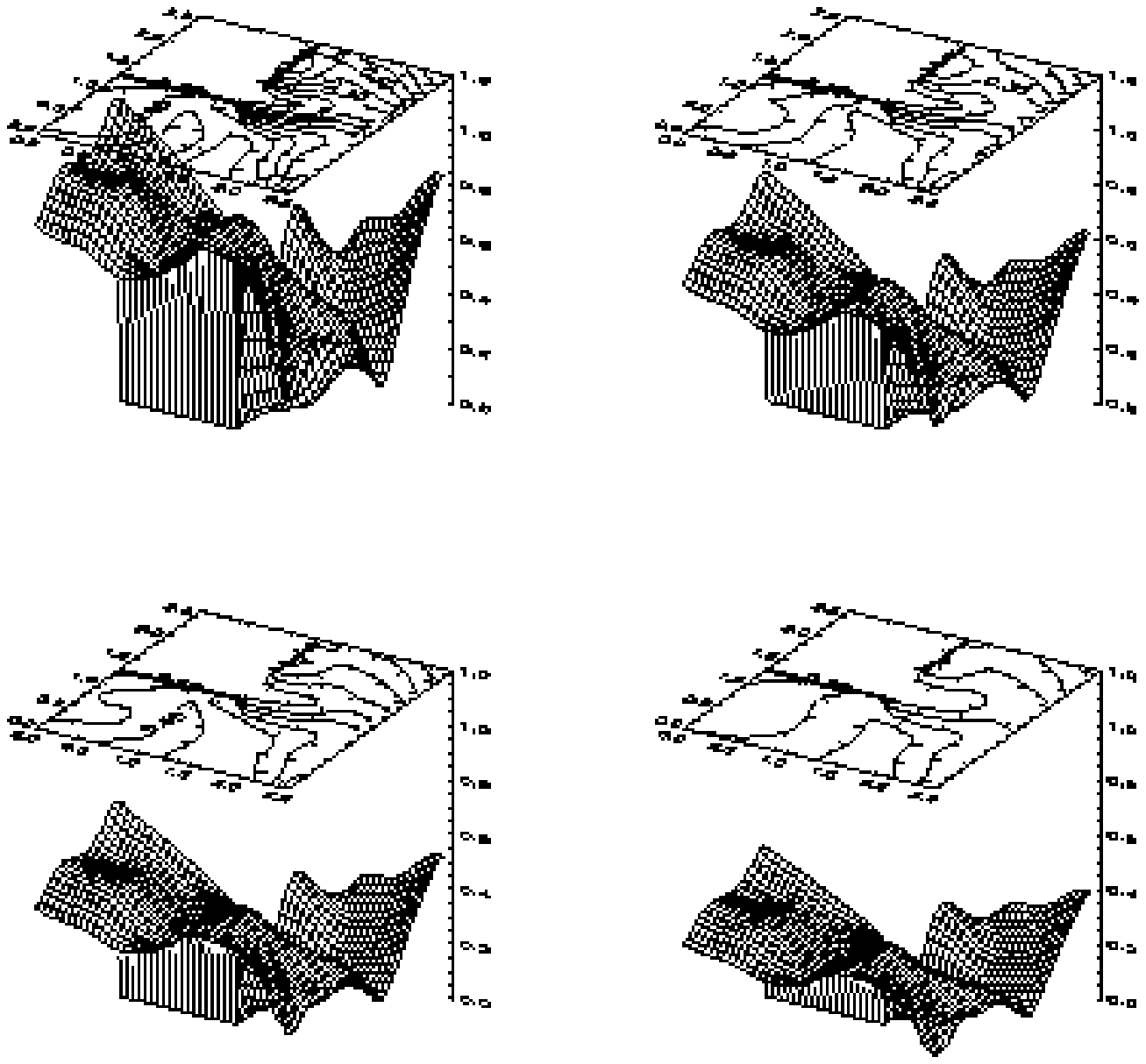}{15}

\xfigure{6}
{The reconstructed surface mass density obtained for the galaxy
samples corresponding to the shear field shown in Fig.\ts 3: $R\in
(24,25.5)$ (upper left), $R\in(23,25.5)$ (upper right),
$R\in(22,25.5)$ (lower left) and $R\in(21,25.5)$ (lower right). For
all four reconstructions we assume that the redshift distribution is
given through (3.8) with $\beta=1$ and $\ave
z=0.8$}{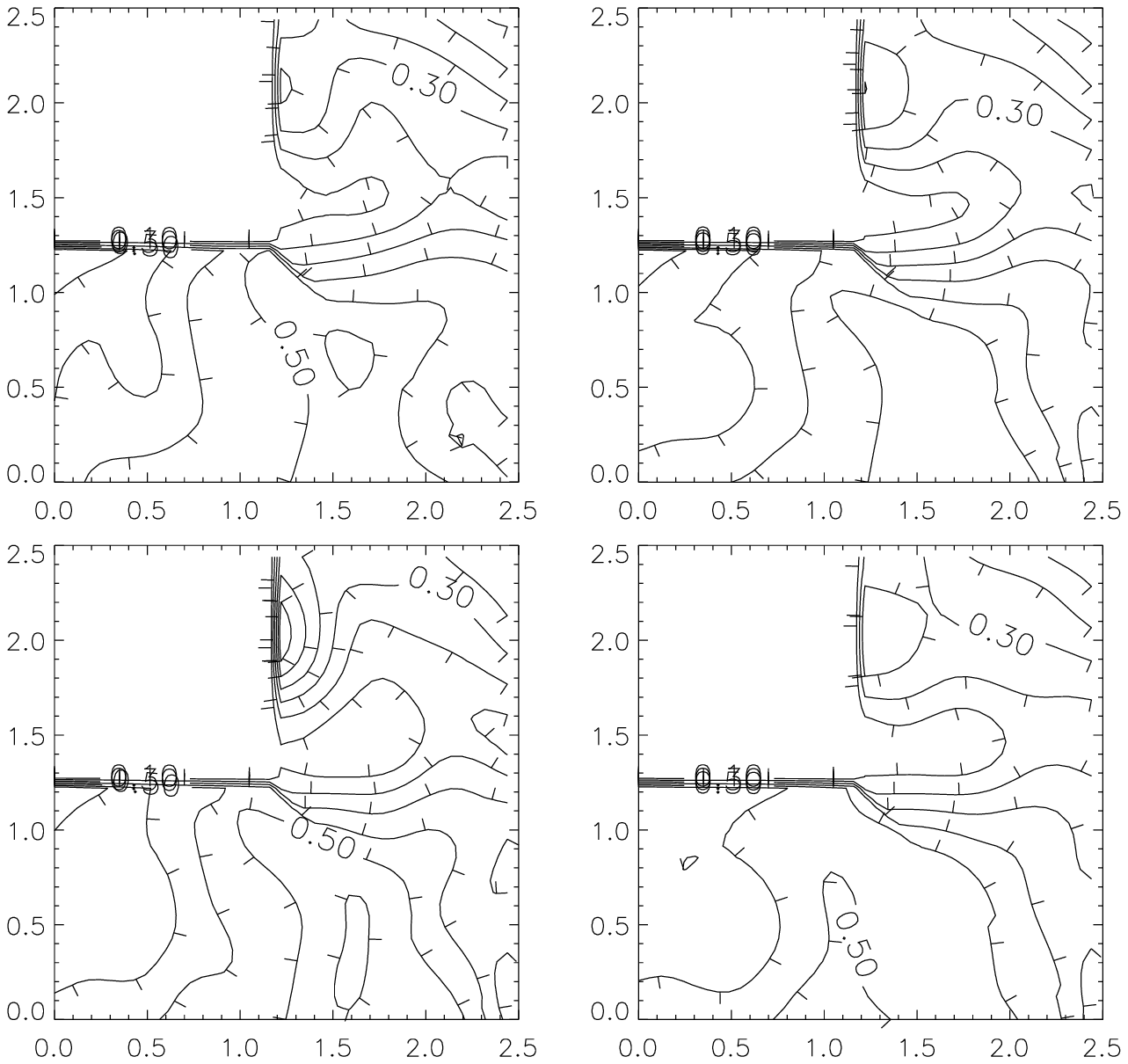}{15}

In Fig.\ts 6 we show the mass density distributions corresponding to
the four shear fields presented in Fig.\ts 3 assuming the redshift
distribution (3.8) with $\beta=1$ and $\ave z=0.8$. As expected from
the shear fields, the mass distribution does not change dramatically. We
find an overall decrease in the mass density for the brighter galaxy
samples from $R\in(23,25.5)$ to $R\in(21,25.5)$. However, the
faintest sample with $R\in (24,25.5)$ gives a mass distribution with a
slightly smaller maximum than that of $R\in(23,25.5)$. We think that
this is not significant and may be due to the fact that fewer images
are used.

\xfigure{7}
{The reconstructed surface mass density for different values of the
smoothing length $s$, with $\ave z=1$ and $\beta=1$. For the upper
left panel we use $s=0 \arcminf 2$, for the upper right $s=0 \arcminf
25$, for the lower left $s=0 \arcminf 35$ and for the lower right $s=0
\arcminf 4$. Note that the main features are common to all mass
distributions shown}
{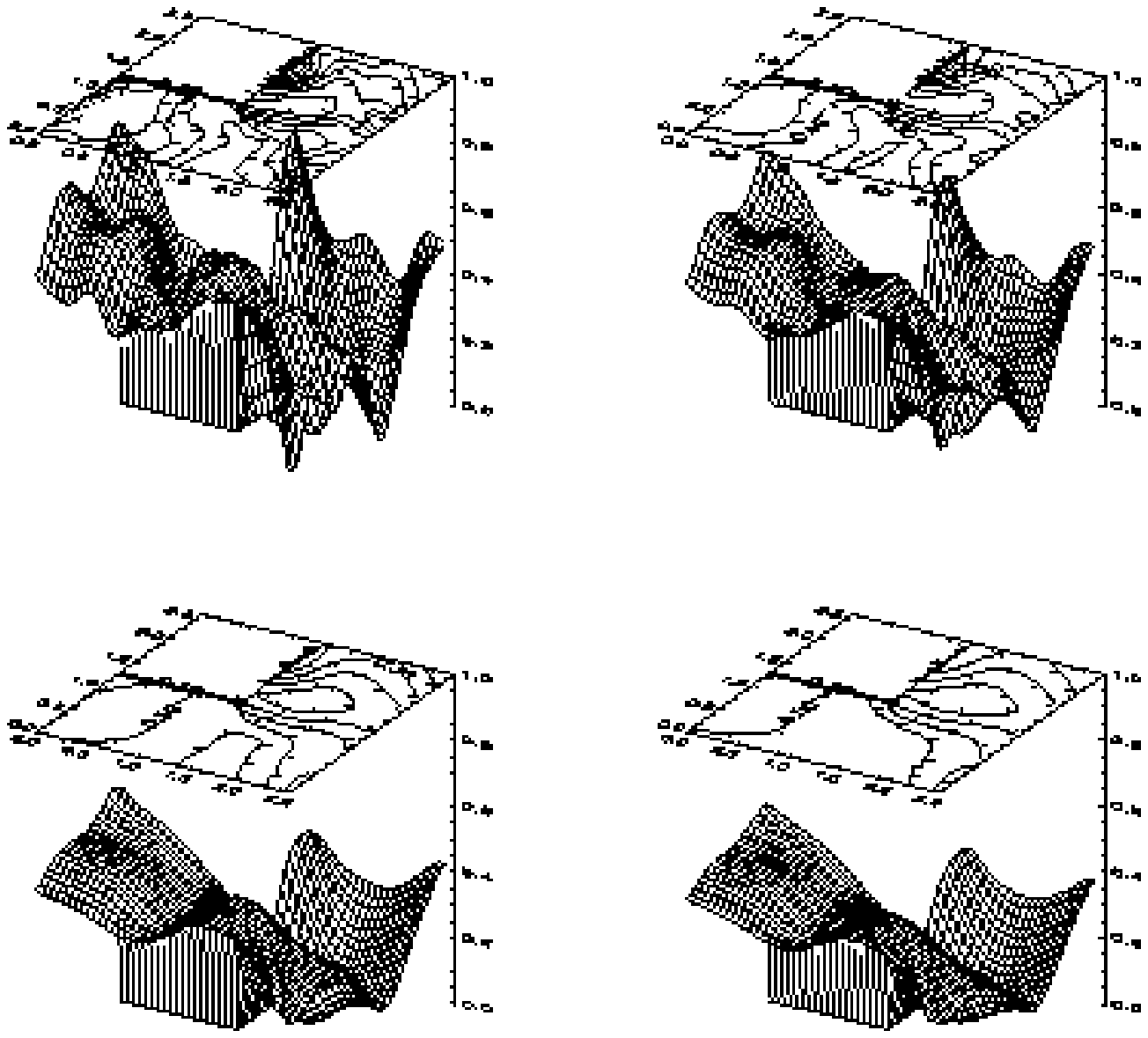}{15}

To investigate the stability of the reconstructed mass distribution,
we repeat the reconstruction for the same parameters $\ave z=1$ and
$\beta=1$ as used for the reconstruction shown in the lower left panel
of Fig.\ts 5 ($s=0 \arcminf 3$), but vary the smoothing length. The
results are shown in Fig.\ts 7 for $s=0
\arcminf 2$ (upper left) , $s=0 \arcminf 25$ (upper
right), $s=0 \arcminf 35$ (lower left) and $s=0 \arcminf 4$ (lower
right). We find that independent of the smoothing length all main
features can be recovered. Obviously, a smoothing length $s=0 \arcminf
2$ gives a too noisy reconstruction, whereas $s=0 \arcminf 4$ may smooth out
too much of the structure. From visual inspection we decieded to use a
smoothing length of $s\approx 0 \arcminf 3$ in the remainder of this
paper. We would like to note that a fixed smoothing length is not
necessarily the best choice, but a smoothing length, adapted to the
local signal strength, may be more appropriate. Such a local adaption
can be objectively controlled with local $\chi^2$-statistics, or by
using regularized maximum-likelihood inversion techniques (Bartelmann
et al.\ts 1995).

As a further check for the stability and reliability of the
reconstructed mass distribution, we perform a bootstrap analysis: we
use the data set consisting of the position-vectors and ellipticities
of the $N_{\rm gal}=295$ faint background galaxies with
$R\in(23,25.5)$ and generate a number of synthetic data sets by
drawing $N_{\rm gal}$ galaxies at a time with replacement from the
original data set. For each of the synthetic data sets we perform the
mass reconstruction. Mass distributions from three different
bootstraps are shown in the upper left, upper right and lower left
panels of Fig.\ts 8. Taking into account that in the bootstrap
analysis on average $1/e\approx 36\%$ of all galaxies are not used at
all, the fact that the main features are still recovered increases our
confidence in the reconstruction.
The average mass density of 30 bootstraps, shown
in the lower right panel of Fig.\ts 8, is very similar to the mass
distribution shown on the lower left of Fig.\ts 5, where all
galaxies and the same smoothing length $s=0 \arcminf 3$ are used.
Comparing the mass reconstructions obtained from different
bootstrapping realizations, one can see that the relative variations
are considerably larger near the boundary of the data field. This is
due to the fact that a point on the boundary has fewer neighboring
galaxies, and thus takes into account less information of the local
shear. We want to stress, however, that this is a `random' noise
components, and not a systemmatic boundary effect.  

\xfigure{8}
{Three mass distributions (upper left and right, lower left) resulting
from different bootstrapping realizations (see text) for $\ave z=1$, 
$\beta=1$ and $s=0 \arcminf 3$. The lower right panel shows the average
of 30 bootstrapping mass distributions}
{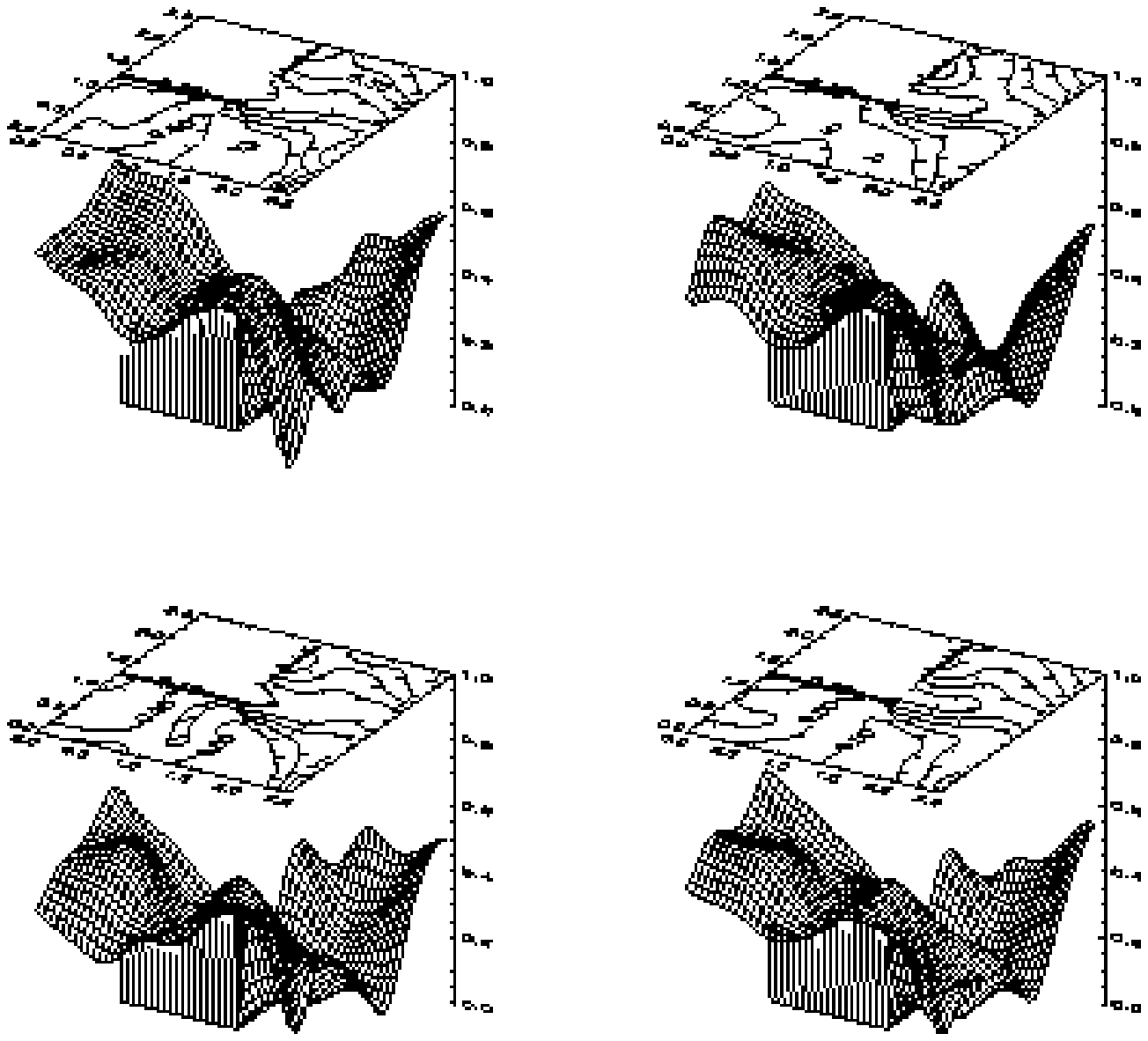}{15}

\subs{4.2 Correlation between mass and light}
We want to compare the reconstructed mass distribution with the light
 distribution of different samples of galaxies. For this we calculate
 the gaussian-smoothed light distribution via
$$
\sigma_{\rm light}(\vc \theta)={\sum_{k=1}
\exp\rund {-{\abs{\vc \theta-\vc \theta_k}^2\over s^2}}\;
10^{-0.4\rund{m_k-20}}\over
 \int_{\cal U} d^2\theta' \exp\rund{-{\abs{\vc
\theta-\vc \theta'}^2\over s^2}}}\ ,
\eqno(4.1)
$$
where $\cal U$ is the data field (i.e. the three quadrants),
$\vc \theta_k$ and $m_k$ are the positions and magnitudes of the
galaxies used, and a smoothing scale of $0 \arcminf 3$ is used.
The denominator in (4.1) corrects for boundary effects.

In Fig.\ts 9a we show the light distribution of all galaxies [roughly
$r\in(17,23)$] detected by Dressler \& Gunn (1992) on a field of
$4\arcmin \times 4 \arcmin$.  Comparing this with the mass
distribution shown in Fig.\ts 5 \& 6 we detect a remarkable
correlation: the position of the maximum in the mass density
corresponds reasonably well with the position of the maximum in the
light distribution, which is approximately located there where
Dressler \& Gunn (1992) proposed the cluster center.  The secondary
mass maximum corresponds to a group of bright galaxies. It is more
prominent in the light than in the mass distribution and displaced
slightly to the left relative to the position of the secondary maximum
in the mass. The minimum of the mass distribution corresponds to a
region where very few bright galaxies are observed.

Dressler et al. (1994b) studied the morphology of the bright (cluster)
galaxies with the HST (WFPC1); we show in Fig.\ts 9b the light
distribution of their identified E/S0 galaxies, tracing the old
cluster galaxy population. The position of the secondary maximum in
this light distribution corresponds better with the position of the
secondary mass maximum and the correspondence with the other features
is as good. Hence we conclude that there is a correlation between the
reconstructed mass 
distribution and the light distribution of the bright galaxies, which
are mostly cluster galaxies.

\xfigure{9}
{(a) The gaussian-smoothed light distribution defined through
Eq.\ts(4.1) of all galaxies [roughly with $r\in(17,23)$] detected by
Dressler \& Gunn (1992) on a field of $4\arcmin\times 4 \arcmin$. We
use a smoothing length $s=0 \arcminf 3$.  (b) The gaussian-smoothed
light distribution of all E/S0 galaxies identified in the field of
about $2 \arcminf 5\times 2
\arcminf 5$ [WFPC1, Dressler et al.\ts (1994b)].
The area covered by the HST (WFPC2) observations is indicated by the
solid lines} {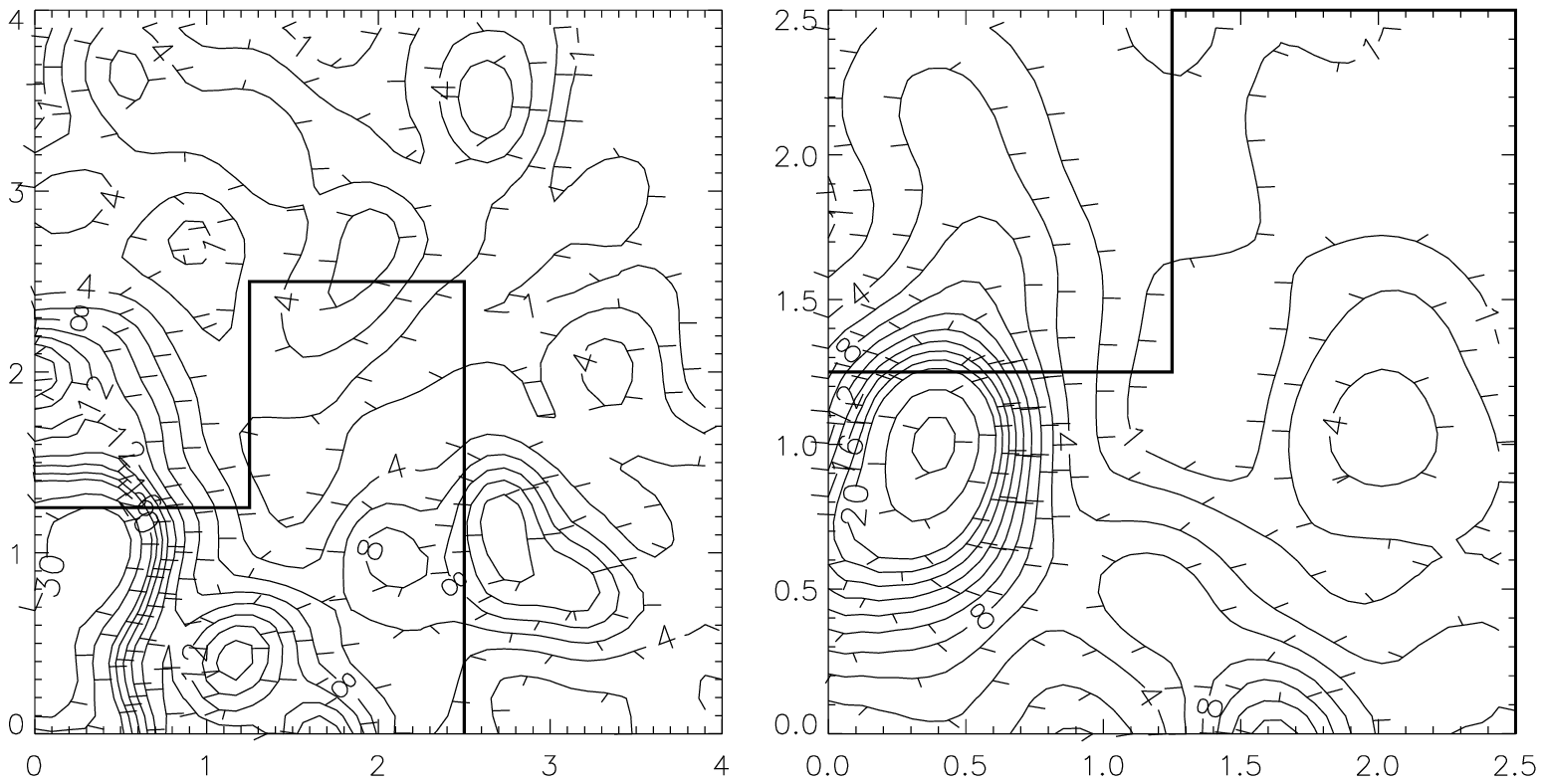}{12}

To investigate the correlation between mass and light more
quantitatively, we calculate from each mass distribution
$\kappa_\infty (\vc \theta)$ obtained in a bootstrap realization (see
Sec.\ts4.1) the number
$$
V:={\sum_{\rm galaxies}\eckk{ \kappa_\infty({\vc \theta}_{\rm galaxy})
-\ave{ \kappa_\infty} } \over N}\ ,
\eqno(4.2)
$$
for different samples of $N$ galaxies, where $\ave{ \kappa_\infty}$ is
the average of $\kappa_\infty$ over our data field ${\cal U}$.
If the galaxies were randomly
distributed, the expectation value of $V$ would be zero, whereas a
positive (negative) correlation of galaxies with the reconstructed
mass density is indicated by $V>0$ $(V<0)$.

\xfigure{10}
{(a): The normalized distribution $p(V)$ of the quantity $V$ defined
in Eq.\ts(4.2), calculated from reconstructed mass distributions for
$1000$ bootstrap data sets drawn from the observed one
($R\in[23,25.5])$ with replacement. The solid curve shows the
distribution for all galaxies detected from the WFPC2 observations,
the dotted curve for all galaxies with $R\in[24,25.5]$, the long
dashed curve for all galaxies detected in the two right CCD frames
with $R\in[24,25]$ and the dashed curve for E/S0 galaxies identified
Dressler et al.\ts (1994b).  (b): The mean correlation coefficient
$\ave V$ for different galaxy samples chosen. The tilde indicates that
the subsample has no well-determined flux threshold}
{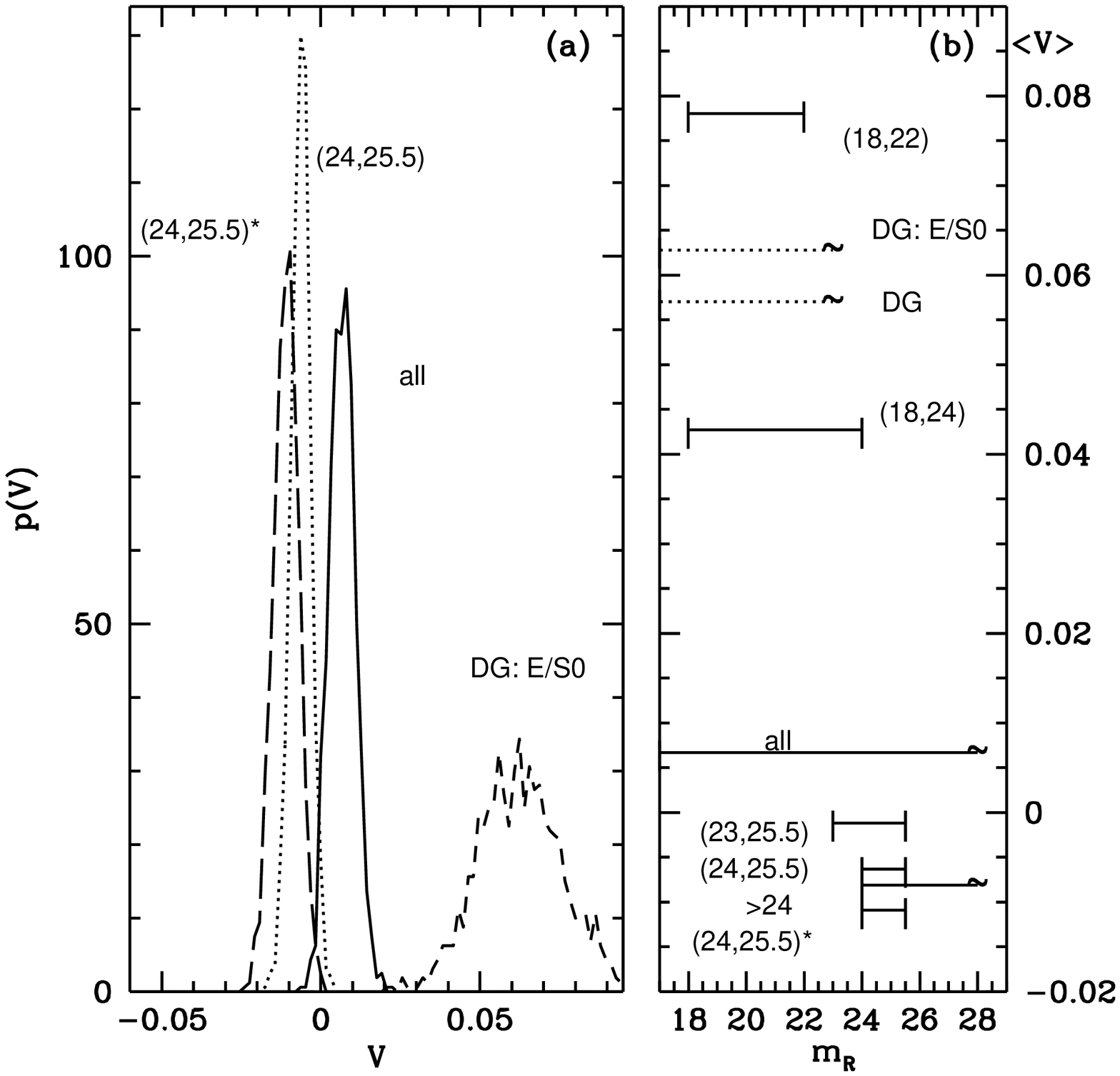}{11}

From $1000$ bootstrap realizations we find the distributions $p(V)$
shown in Fig.10a for the E/S0 galaxies identified by Dressler et
al.\ts (1994b) (DG: E/S0: dashed curve), all galaxies detected from
the WFPC2 data regardless from their size (solid curve) and all
galaxies with $R\in(24,25.5)$ (dotted curve) -- also regardless of
size.  Clearly, a strong positive correlation between the
reconstructed mass and the DG:E/S0 galaxies is detected. Next, a
weaker positive correlation between `all' galaxies and the mass
distribution, and an anti-correlation between the mass distribution
and the faint galaxies [$R\in(24,25.5)$] is visible. This
anti-correlation appears surprising on first sight, as certainly some
of the faint galaxies belong to the cluster, and, as we have argued
above, the cluster galaxies are positively correlated with the
mass. To investigate this point further, we have calculted the
distribution of $V$ for the same magnitude interval, but leaving out
the lower left CCD where the contribution from cluster galaxies is
expected to be strongest. The resulting distribution is also plotted
in Fig.\ts 10a, indicated with an asterisk; it shown an even stronger
anti-correlation.

In Fig.\ts 10b we show the mean correlation coefficient $\ave V$ from
$1000$ simulations as a function of the magnitude range of the
galaxies. We find that $\ave V$ is strongly correlated with the
faintness of the magnitude interval $(m_1,m_2)$ chosen. It decreases
towards the fainter samples and eventually becomes negative.  This is
due to the larger fraction of background galaxies contributing to the
counts in $(m_1,m_2)$ for fainter slices.

We now turn to a possible explanation for the anti-correlation of the
faint galaxies with the reconstructed surface mass density:

The locally observed number
density $n_L(>S)$ of lensed background galaxies with flux larger than
$S$ is related the unlensed number density
$n_0(>S)$ through the local magnification $\mu$ caused by the
cluster,
$$
n_L(>S,\vc \theta)={1\over \mu(\vc \theta)} \;  n_0\rund{ >{S\over \mu(\vc
\theta)}, \vc \theta}  \ ,
\eqno(4.3)
$$
where
$$
\mu(\vc\theta)=\int_0^\infty dz\; p_s(z)\;{1\over
\abs{[1-w(z)\kappa_\infty(\vc\theta)]^2- 
w(z)^2\gamma_\infty^2(\vc\theta)}}\ ,
\eqno(4.4)
$$
is the redshift-averaged local magnification, weighted by the redshift
distribution of the galaxies.  The first factor in (4.3) is due to the
increase of the solid angle, whereas the argument of $n_0$ indicates
that a magnified source can be `intrinsically' fainter by a factor
$\mu$ and still be included in a flux-limited sample.  Which of the
two competing processes wins depends on the slope $n_0(>S)$ of the
sources. The observed galaxy counts in the R-band shows that it is
well fitted by $N(R)\propto 10^{\gamma R}$ with
$\gamma=0.32$. Therefore, we obtain from Eq\ts.(4.3)
$$
{n_L(>S)\over n_0(>S)}=\mu^{2.5\gamma-1}\ .
\eqno(4.5)
$$
The magnification (see also Broadhurst, Taylor \& Peacock
1995, hereafter BTP) can be obtained via
$$
\mu=\eck{{n_L(>S)\over n_0(>S)}}^\alpha \qquad {\rm with}\qquad 
\alpha={1\over 2.5\gamma-1} \ .
\eqno(4.6)
$$
For $\gamma=0.32$, the exponent in Eq.\ts(4.6) is $\alpha=-5$ and a
 suppression of background galaxies is expected in regions of high
magnifications, or high surface mass density. 

In Fig.\ts 11 we show the gaussian-smoothed number density [defined as
in (4.1) without flux weighting] of the faint galaxies with
$R\in(24,25.5)$. We see a local maximum where we detect the minimum of
the mass, indicating that we found the expected
anti-correlation. However, we also find two local maxima in the number
density of the faint objects where we detect the maxima of the
mass. Note that we have not corrected the faint galaxy density for
occupation of some CCD area by bright (cluster) galaxies, which is of
course strongest near the cluster center; this shows that the
contribution of cluster members to the faint galaxy counts is slightly
stronger than indicated by comparing Fig.\ts11 with Fig.\ts9.  We thus
conclude that a non-negligible fraction of the faint galaxies are
cluster members, as also follows from Fig.\ts 2.

\xfigure{11}
{The gaussian-smoothed number density
of the faint galaxies with $R\in (24,25.5)$. We use a smoothing length
$s=0 \arcminf 3$}{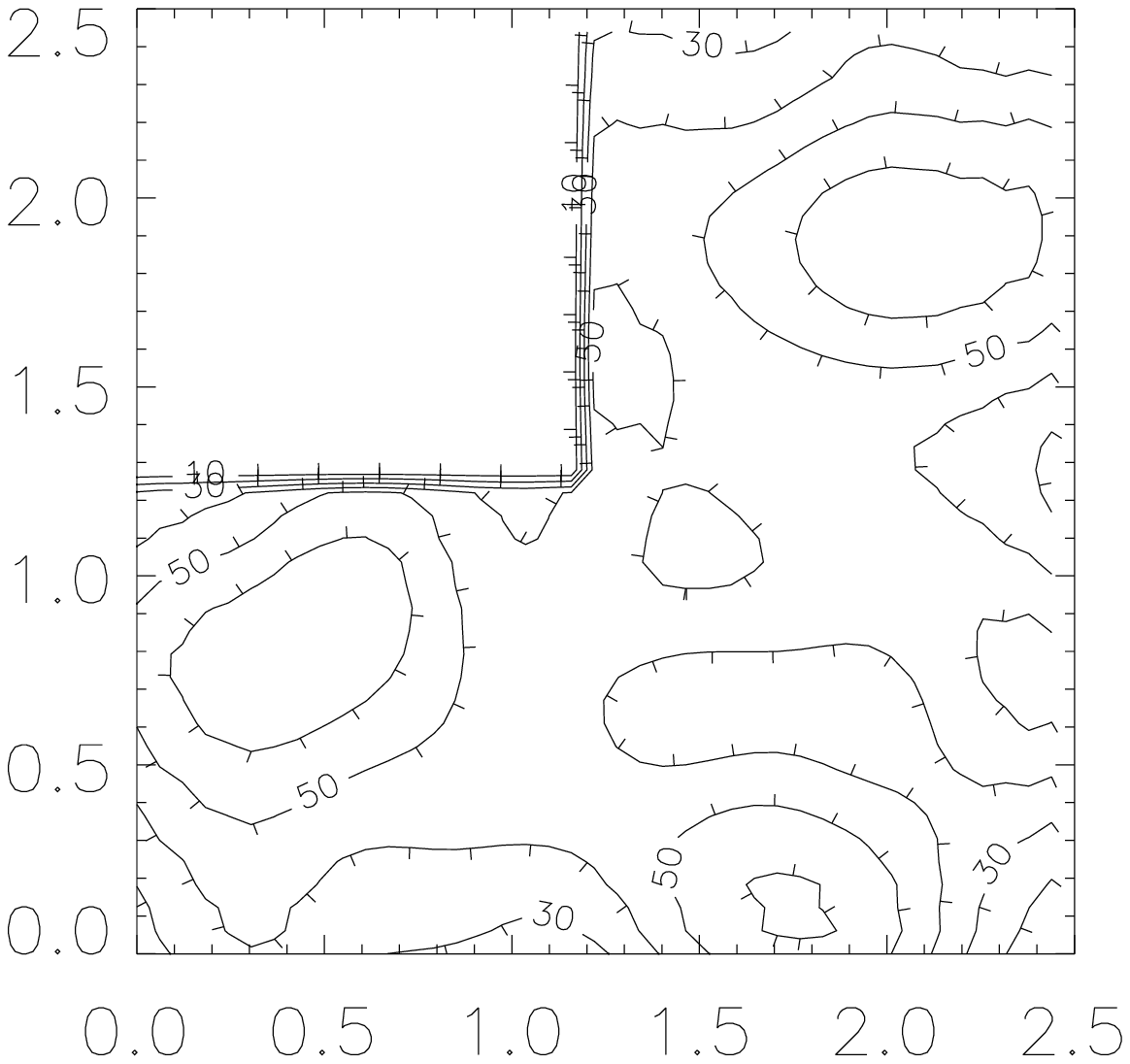}{6}

Assuming that the cluster galaxies have an average correlation
coefficient $\ave V_c>0$ with the mass distribution, independent of
the magnitude of the galaxies, and that the background galaxies have an
average correlation coefficient $\ave V_b<0$, again independent of
their magnitudes, one can then derive the fraction $x$ of cluster
galaxies in a magnitude slice $(m_1,m_2)$ through measuring
$$
\ave V_{m_1,m_2}= x \ave V_c+(1-x) \ave V_b \ .
$$
Using $\ave V=\ave V_c$ from the galaxies with $R\in(18,22)$ and $\ave
V_b$ from the galaxies with $R\in(24,25.5)* $ we estimate the fraction of
cluster galaxies to be $83\%$ for the `DG E/S0' sample, $76\%$ for the
DG sample, $60\% $ for the galaxies with $R\in(18,24)$, $11\%$ for
$R\in(23,25.5)$ and $5\% $ for $R\in(24,25.5)$. Of course these values
are crude estimates only, but they do not appear unreasonable.

To summary this subsection, the correlation of bright galaxies with
the reconstructed surface mass density shows that {\it in this cluster
`mass follows light' on average.} Hence, {\it overdensities of bright
galaxies correspond to local maxima in the projected mass density.}
The also significant anti-correlation of faint galaxies with the
reconstructed mass profile is most likely an effect of the
magnification (anti)bias, which has been pointed out by Broadhurst,
Taylor \& Peacock (1995) and which was detected in the cluster A1689
(Broadhurst 1995).

\subs{4.3 Limits on the mass inside the data field}
Requiring that the surface mass density can not be negative one can
obtain a lower limit on the mass inside the data field by applying the
invariance transformation (3.5) such that the minimum of the resulting
$\kappa_\infty$-map is zero.  Using the galaxies with $R\in(23,25.5)$,
we find as a {\it lower bound} on the total mass inside the data field
(side length about $1$ Mpc $h^{-1}_{50}$) about $M/(10^{14}h_{50}^{-1}
M_\odot)\ge 6.3 $ ($4.3,3.6,2.8 $) for a mean redshift
$\ave z=0.6$ ($0.8,1.0,1.5$). These limits depend only slightly on the
actual form of the assumed redshift distribution, as shown in Fig.12.
\xfigure{12}{Lower limits $M_{\rm min}$ 
on the total mass inside the data field in units of
$M_{14}=10^{14}h_{50}^{-1} M_\odot$ as a function of the assumed mean redshift
$\ave z$ (left) or the mean $\ave w$ (right) of the
images used for the reconstruction, assuming the redshift distribution
(3.8).
Crosses show the results for
$\beta=1$, triangles for $\beta=1.5$ and dashes for $\beta=3$. The
smoothing length is $s=0 \arcminf 3$ }
{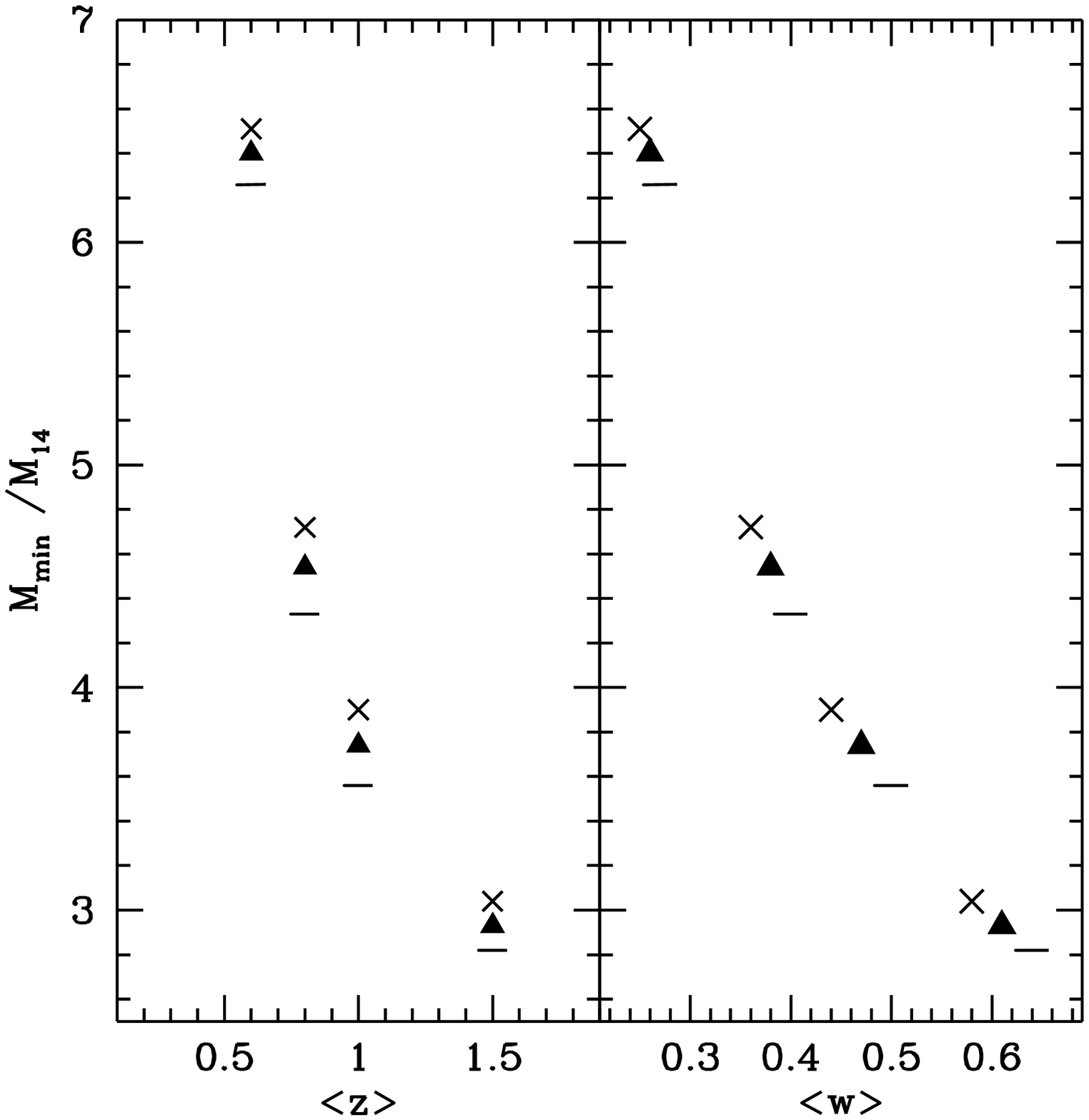}{9}

The most conservative {\it upper limit} of the mass we can give is
$M<10 ^{15} h^{-1}_{50}M_\odot$, because this corresponds to
$\ave{\kappa_\infty}=1$ and would most probably produce several giant
arcs which are not observed.
 
Using Eq.\ts(4.6), one can in principle derive the local magnification
$\mu(\vc \theta)$ and could therefore break the global invariance
transformation (3.5) with the measurement of $\mu(\vc \theta)$ at one
particular point in the cluster.  However, in practice we have the
following difficulties: (1) since lensing effects are nowhere weak on
the whole field we have no measurement of $n_0(>S)$; (2) we have no
colour information of the galaxies and therefore we can not well
distinguish between faint cluster galaxies and lensed background
galaxies; (3) possible clustering of background objects and confusion with
cluster member galaxies leads to a high noise in the estimate of
$\mu(\vc \theta)$ and a poor resolution.  Therefore, we can not derive
an accurate high-resolution map of the magnification from Eq.\ts(4.6).

Nevertheless, we can use the magnification (anti) bias to derive
(crude) estimates of the total mass inside the data field. Crucial for
this is the assumption that the unlensed number counts $n_0(>S)$ can
be taken from the literature. We use the amplitude and slope given in
Smail et al. (1995b) which gives for the data field $\cal U$ and
galaxies with $R\in(23,25.5)$
$N_0(23,25.5)=278$. The observed number of $N^{\rm
obs}(23,25.5)=N^{FG}+N^{CG}=295$ is a sum of cluster galaxies (CG) and
galaxies belonging to the faint galaxy field population (FG). Now, we
assume that a fraction $x$ of the observed galaxies $N^{\rm obs}$ are
cluster galaxies and obtain
$$
\ave{\mu^{2.5\gamma-1}}_{\cal U}= {N^{FG}\over N_0}={(1-x)N^{\rm obs}
\over N_0} \ .
\eqno(4.6)
$$
For $x=0.15 $ ($0.2$) we obtain $\ave{\mu^{2.5\gamma-1}}_{\cal
U}=0.849$ ($0.796$). We perform the reconstruction assuming a value of
$\lambda$, or equivalently, a value of $\ave{\kappa_\infty}$, 
and calculate from the resulting mass map $\kappa_\infty $ and the
shear map $\gamma_\infty$ locally the expectation value of the
magnification (4.4) of the sources
and finally average $\mu^{2.5\gamma-1}$ on the field $\cal U$ to
obtain $\ave{\mu^{2.5\gamma-1}}_{\cal U}$.  Next, we search for that
value of $\ave {\kappa_\infty}$ which gives the value
$\ave{\mu^{2.5\gamma-1}}_{\cal U}$ corresponding to a certain fraction
$x$ of cluster galaxies. In Fig.\ts 13 we show the mass estimates for
$x=0.15$ (solid curve) and $x=0.2$ (dotted curve) as a function of the
assumed mean redshift of the galaxies used for the reconstruction. For
comparison we show again the minimum mass (dashed curve) as shown in
Fig.\ts 12 for $\beta=1$.  From these lower limits on the mass we
conclude that the fraction of cluster galaxies in the galaxy sample
with $R\in(23,25.5)$ is $x\gtrsim 0.1$. 

\xfigure{13}{The mass $M$ inside the data field in units of
$M_{14}=h^{-1}_{50} 10^{14} M_\odot$ as a function of the assumed mean
redshift $\ave z$ (left) or the mean $\ave w$ (right) of the images
used for the reconstruction, assuming the redshift distribution (3.8)
with $\beta=1$. The solid curve shows the total mass in the field
$\cal U$ assuming that $x=15\%$ of all galaxies detected within
$R\in(23,25.5)$ are cluster galaxies, the dotted curve shows the
result for $x=20\%$.  For comparison we show $M_{\rm min}$ from
Fig.\ts 12 (dashed curve)}
{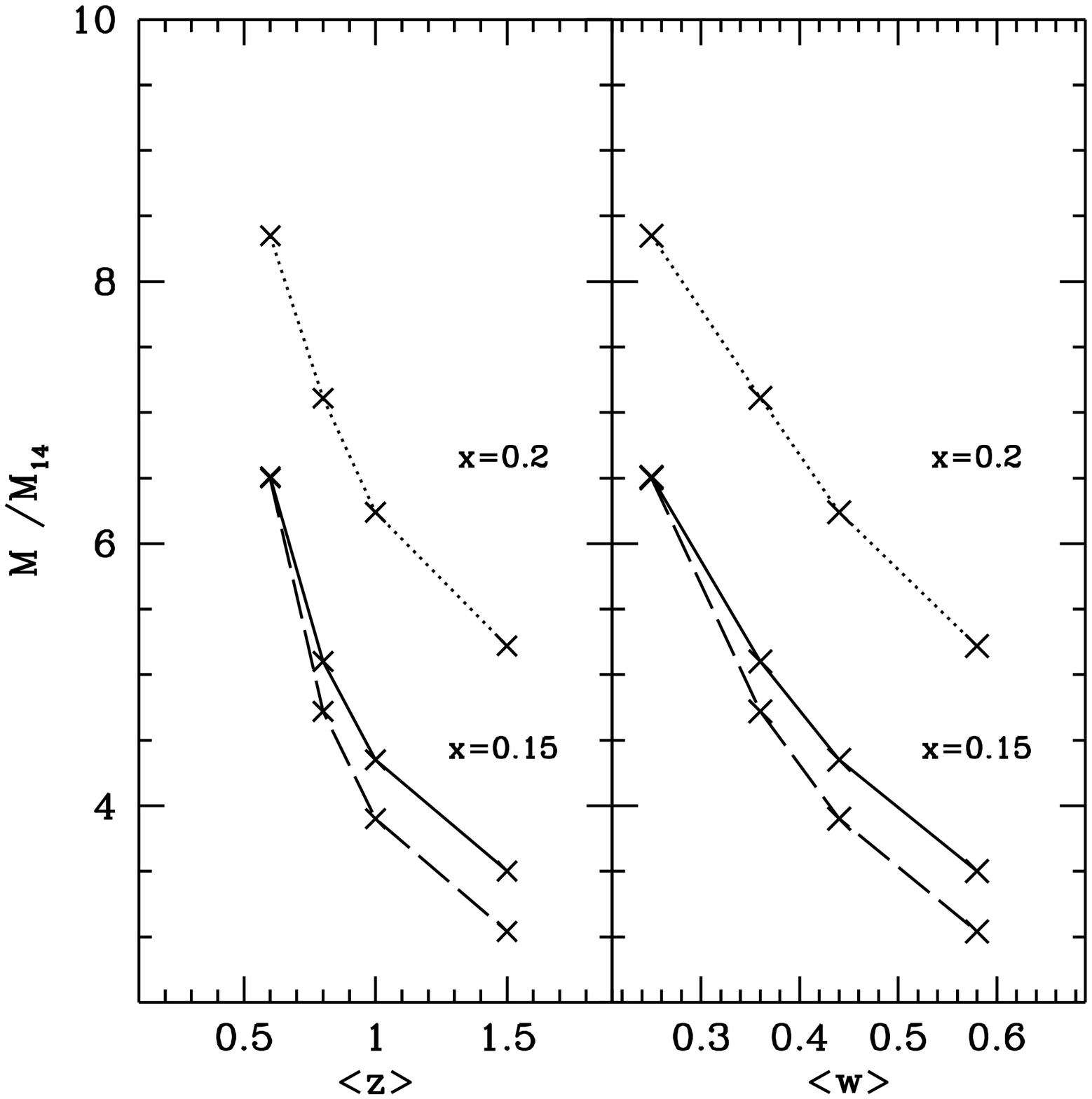}{9}
We note that  recently we became aware of a paper by Belloni et
al. (1995) which could allow for a 
better
separation between cluster and background galaxies for the galaxies
with $R\in(23,25.5)$: Belloni et al. used multiband
photometry to get the redshifts of 275 bright galaxies with $R<22.5$,
i.e., the fraction of cluster galaxies in the sample of galaxies  with
$R<22.5$. From that and the known slope of the faint field galaxies,
we can probably derive a better estimate for the fraction of cluster
galaxies in the faint galaxy sample with $R\in(23,25.5)$. 
\subs{4.4 The mass to light ratio}
We calculate the total light of all galaxies inside the field $\cal U$
detected by Dressler et al. (1994b) leaving aside those galaxies whose
measured redshift exclude a cluster membership. Magnitudes for these
galaxies are given in gunn-r ($\bar \lambda=655$ nm), which correspond
to a rest-frame wavelength of $\lambda \approx 464$ nm), i.e. the
measured $r$ magnitudes correspond to the $B$ ($\bar \lambda=443$
nm) magnitudes in the rest-frame. 
As a result we obtain for this sample a total luminosity of
$(L/L_\odot)_B=5 \times 10^{12} h^{-2}_{50}$.  From this and the mass
estimates shown in Fig.\ts 13 we derive the $M/L$ values shown in
Fig.\ts 14.  If the Dressler et al. (1994b) sample of galaxies with
$r\in(17,23)$ represents the luminosity of cluster galaxies well
and if the mean redshift of the faint galaxies [$R\in(23,25.5)$] used
for the mass reconstruction is $\ave z=0.8$, then we derive from the
non-negativity of the surface mass density a lower limit on
$M/L\gtrsim 93h_{50} $ (dashed curve in Fig.\ts14) for the cluster
CL0939+4713, and the values $M/L\approx 102h_{50} $ for a fraction of
$x=0.1$ of cluster galaxies in the faint galaxy sample (solid curve)
and $M/L\approx 142 h_{50}$ for $x=0.2$ (dotted curve). From the
absence of giant luminous arcs we derive - independently of the
assumed redshift of the sources or the fraction of cluster galaxies -
a robust upper limit $M/L \lesssim 200 h_{50}$. Of course, the sample
chosen for calculating the luminosity of the cluster includes a
certain fraction of background- and foreground galaxies, but this is
partly compensated because we will miss
a certain fraction of faint cluster galaxies. To derive a conservative
upper limit of the total luminosity of the cluster, or an conservative
lower limit on $M/L$ we calculate the total luminosity of all galaxies
detected in the field (down to $R=26.5$) and obtain almost twice the
value $(L/L_\odot)_B$ found above, or half the values for $M/L$ as
shown in Fig.\ts14.

\xfigure{14}
{The values of $(M/L)_B$ in units of $h_{50} (M_\odot /L_\odot)_B$ 
obtained for the masses shown in Fig.\ts 13. To
calculate the luminosity $L$ we used all galaxies identified by
Dressler \& Gunn (1992) in that field, excluding those galaxies for
which a measured  redshift shows that they are no cluster members}
{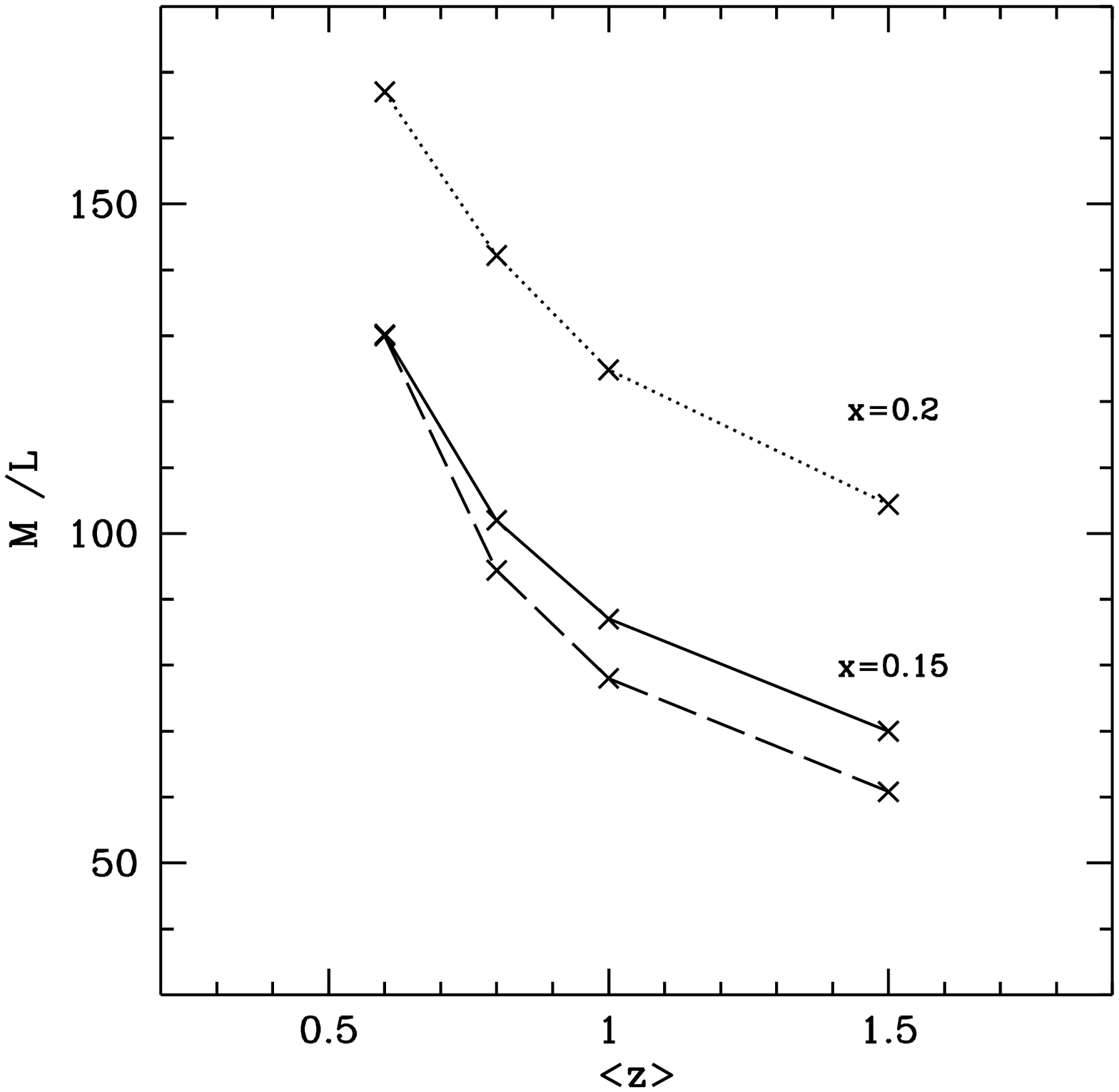}{8}

The fact that the $M/L$ values (see Fig.\ts 14) 
are small compared to the $M/L$ found
for other cluster from a weak lensing analysis 
(see Fahlman et al. (1994), or Smail et al. (1995a)),
 is not too surprising taking into
account that CL0939+4713 is an {\it optically} selected (Abell) cluster with 
high redshift $z_d=0.41$. A moderately bright cluster at this redshift
would probably not have entered the Abell catalog. In addition, the
$M/L$-ratio quoted here is uncorrected for cosmic evolution, which can
be quite substantial from $z=0.4$ to today in the B-band, so that the
corresponding $M/L$ value `today' would be considerably higher.

\subs{4.5 The Rosat PSPC image}
In Fig.\ts 15 we show the Rosat PSPC image of the cluster obtained
from a gaussian smoothing of the counts with a smoothing length of $s=0 \arcminf 2$.
The center of the coordinate frame coincides with the lower left corner
of the HST image shown in Fig.\ts 1. To align the PSPC image with the
HST image we used a star. Thus, the
alignment slould be better than one PSPC pixel (pixelsize
$15\arcsec$). The PSPC image shows a main X-ray emission around the cluster
center, which roughly corresponds to the region where we derive a
maximum of the light and a maximum of the mass distribution. 
The PSPC data are analysed in a forthcoming paper by S. Schindler.
In order to perform a more detailed comparison between the X-ray and
mass distribution one has to wait for the ROSAT HRI image. 

\begfig 9 cm
\figure{15}
{The PSPC image of the cluster CL0939+4713. The contour-lines show the
photon counts, ranging from 9.5 down to 1.5 with a spacing of 1. North
is at the bottom and east to the right}
\endfig
 
\sec{5 Discussion}
Using deep WFPC2 data, we have reconstructed the projected (dark)
matter distribution of the cluster Cl0939+4713. The
distortion of faint background galaxies was used to construct a `shear
map' of the cluster, from which an unbiased, nonlinear estimate of the
surface mass density was constructed. The resulting mass map is
defined up to an overall invariance transformation, a generalization
of the so-called mass sheet degeneracy. The mass distribution is
strongly correlated with the projected distribution of the bright
cluster galaxies; in particular, the maximum of the mass map coincides
with the cluster center as determined from the light distribution, a
secondary maximum of the map corresponds to a concentration of cluster
galaxies, and a deep mass minimum occurs where the number density of
cluster galaxies is lowest. We also note that the main mass (and light)
maximum correspond to maximum in the X-ray emission, as seen
with the ROSAT PSPC. The
anti-correlation of mass with faint galaxies is interpreted as the
difference between a positive correlation of mass with faint cluster
galaxies (mainly seen towards the cluster center), and a magnification
anti-bias (BTP), which is expected due to the flatness of the galaxy number
counts. 

Our analysis shows that the recently developed cluster inversion
techniques can be applied to (sufficiently deep)  WFPC2 data (in order to
image precisely faint galaxies), despite the fact that its
field-of-view is fairly limited. It is {\it essential} to use an
unbiased finite-field inversion technique in this case, and also,
since the cluster center is (nearly) critical, to account for strong
lensing effects. Also, owing to the fairly large redshift of the
lensing cluster, the redshift distribution of the background galaxies
has to be taken into account explicitly; only in weak lensing regime
does this distribution not enter the reconstruction, but only the mean
of the distance ratio $D_{ds}/D_s$. 

We have checked the robustness of the mass reconstruction, by using
different magnitude cuts for the galaxies and by extended bootstrap
simulations. {\it The main features of the mass map -- the two mass
maxima, the pronounced minimum, and the overall gradient toward the
cluster center -- are stable.} The anti-correlation of mass with the
faint galaxies, and the strong correlation with cluster galaxies,
further increases our confidence in the reconstruction. We have
derived estimates of the cluster mass contained within the WFC
aperture, which depend on the assumed redshift of the background
galaxies. A robust lower limit of the mass follows from the
non-negativity of the surface mass density, a robust upper limit comes
from the absence of giant luminous arcs. To derive a narrower mass
range, one needs to fix the parameter $\lambda$ contained in the
invariance transformation (3.5). This can be done by using the
magnification anti-bias (BTP). In the only case where this effect has
been demonstrated before (A1689, Broadhurst 1995), a color criterium was used
to ensure that the galaxies are likely background galaxies. Since we
lack color information, the fraction of the faint galaxies which are cluster
members or foreground galaxies cannot be separated from background
galaxies. Nevertheless, a plausible range for $\lambda$ can be
obtained and led to the mass estimates shown in Fig.\ts 13.

The exploration of this novel method to reconstruct the density
distribution of clusters has only just begun. In contrast to current
ground-based data, for which image ellipticities have to be
substantially corrected for seeing effects, WFPC2 data provide
relatively `clean' probes of image ellipticities. The small field of
view of WFPC2 limits the extent to which clusters can be mapped
(specially for nearby clusters), unless mosaics are taken, but
high-resolution mass maps such as the one constructed here are
invaluable tools for investigating substructure in cluster mass
distributions and their relation to substructure in the distribution of
galaxies and X-ray emission.

{\it Acknowledgement.} CS would like to thank Hans-Walter Rix
for many useful discussion and for introducing her to the IRAF
software package and Sabine Schindler for help
concerning the PSPC data. We gratefully acknowledge enthusiastic discussion
with Ian Smail, Richard Ellis, Yannick Mellier and Bernard Fort on 
cluster-lenses. \SFB
JPK acknowledges support from a HCM-EU fellowship and the hospitality
of the MPA during a visit when this project was started.

\sec{References}
\def\ref#1{\vskip5pt\noindent\hangindent=40pt\hangafter=1 {#1}\par}
\ref{Bartelmann, M. \ 1995, A\&A, 303, 643.}
\ref{Bartelmann, M. \& Narayan, R. \ 1995, ApJ, 451, 60.}
\ref{Bartelmann, M., Narayan, R., Seitz, S. \& Schneider, P. \ 1995,
submitted to ApJ.}
\ref{Bertin \& Arnouts \ 1995, A\&A submitted.}
\ref{Belloni, P., Bruzual, A.G., Thimm, G.J. \& R\"oser, H.J. \ 1995,
297, 61}
\ref{Bonnet, H., Mellier, Y. \& Fort, B.\  1994, ApJ 427, L83.}
\ref{Brainerd, T.G., Blandford, R. \& Smail, I. \ 1995, preprint.}
\ref{Broadhurst, T.J., Taylor, A.N. \& Peacock, J.A. \ 1995, ApJ 438,
49.}
\ref{Broadhurst, T.G. \ 1995, preprint.}
\ref{Coless, M., Ellis, R.S., Broadhurst, T.J., Taylor, K. \& Peterson,
B. \ 1993, MNRAS 261, 19.}
\ref{Cowie, L.L., Hu, E.M. \& Songaila, A. \ 1995, accepted for Nature.}
\ref{Dressler, A. \& Gunn, J.E. \ 1992, ApJSS, 78, 1.} 
\ref{Dressler, A., Oemler, A., Sparks, W.B. \& Lucas, R.A.\  1994a, ApJ
435, L23.}
\ref{Dressler, A., Oemler, A., Butcher, H. \& Gunn, J.E.\  1994b, ApJ
430, 107.}
\ref{Fahlman, G.G., Kaiser, N., Squires, G. \& Woods, D.\  1994, ApJ 437, 63.}
\ref{Fort, B. \& Mellier, Y. \ 1994, A\&AR 5, 239.}
\ref{Gorenstein, M.V., Falco, E.E. \& Shapiro, L.I. \ 1988,ApJ 327, 693. }
\ref{Holtzman, J.A., Burrows, C.J. , Casterno, S., Hester, J.J., Trauger, J.T.,
Watson, A.M., Worthey, G., 1995, preprint.}
\ref{Kassiola, A., Kovner, I., Fort, B. \& Mellier, Y. \ 1994, ApJ 429, L9.}
\ref{Kaiser, N. \& Squires, G.\ 1993, ApJ 404, 441 (KS).}
\ref{Kaiser, N. \ 1995, ApJ 493, L1.}
\ref{Kaiser,N., Squires, G., Fahlman, G.G., Woods, D. \& Braodhurst,
T. \ 1994, preprint astroph/9411029.}
\ref{Kneib, J.P., Ellis, R.S., Smail, I.R., Couch, W.J., Sharples, R., 1995,
ApJ submitted.}
\ref{Kochanek, C.S.\ 1990, MNRAS 247, 135.} 
\ref{Lilly, S. \ 1993, ApJ 411, 501.}
\ref{Miralda-Escude, J.\ 1991, ApJ 370, 1.}
\ref{Schneider, P. \ 1995, A\&A, in press.}
\ref{Schneider, P. \& Seitz, C.\  1995, A\&A 294, 411.}
\ref{Smail, I., Ellis, R.S., Fitchett, M.J. \ 1995a, MNRAS 273, 277.}
\ref{Smail, I., Hogg, D.W., Yan, L. \& Cohen, J.G. \ 1995b, ApJ, 449, L105.}
\ref{Seitz, C. \& Schneider, P.\  1995a, A\&A 297, 287.}
\ref{Seitz, C. \& Schneider, P.\  1995c, in preparation.}
\ref{Seitz, S. \& Schneider, P.\  1995b, A\&A, in press.}
\ref{Tyson, J.A., Valdes, F. \& Wenk, R.A.\  1990, ApJ 349, L1.}

\end